\documentclass[a4paper,11pt]{article}
\pdfoutput=1
\usepackage{jheppub}
\usepackage[T1]{fontenc}
\usepackage{caption}
\usepackage{subcaption}
\usepackage{varwidth}
\usepackage{float}
\usepackage{amsmath}
\usepackage{wrapfig}
\usepackage{amssymb}
\usepackage{framed}
\usepackage{indentfirst}
\usepackage{appendix}

\title{\boldmath Spontaneous scale symmetry breaking at high temperature}

\author{Z. Lalak}
\author{and P. Michalak}
\affiliation{Institute of Theoretical Physics, Faculty of Physics,
\\University of Warsaw, ul. Pasteura 5, 02-093 Warsaw, Poland}

\emailAdd{z.lalak@uw.edu.pl}
\emailAdd{pmichalak@fuw.edu.pl}

\abstract{
We consider a scale symmetric extension of the Standard Model Higgs scalar sector. The new sector, dilaton, is responsible for the generation of mass scales and 
may have geometric origin in the Weyl gravity $\Tilde{R}^2$ term. We show how temperature as a mass scale breaks scale symmetry explicitly and through a nonvanishing thermal vev. In addition, we demonstrate that cosmological evolution of the dilaton-Higgs system can lead to  
late time mass scales, which agree with the Standard Model.
}

\begin{document} 
\maketitle
\flushbottom

\section{Introduction}

The Standard Model (SM) has a number of yet unresolved problems that may indicate that it is a part of a larger and more fundamental theory. This theory would explain the gaps in the current models. One such problem is the so-called hierarchy problem. Despite the fact that SM is a renormalizable theory, to keep the Higgs mass hierarchically lower than the Planck mass scale $M_P$ in the presence of quantum corrections, as 
measured in experiments, one needs to arrange for very precise cancellations between a priori unrelated contributions 
to  
the effective potential of the scalar sector. This problem can be avoided in theory with the mechanism forcing $m_H=0$, i.e. in the model with scale symmetry or conformal symmetry. However, non-zero
Higgs mass must somehow be generated in these types of models. Mechanism generating $m_H$ may be associated with an explicit violation of scale symmetry, e.g. by a quantum anomaly, or with spontaneous breaking of scale symmetry. For consistency, such a theory needs to be coupled with a version of gravity, which, to some extent, respects the scale symmetry. The problem of explaining the Higgs mass hierarchically smaller than the Planck scale is one of the greatest challenges of modern theory of fundamental interactions, therefore all attempts to solve it are in the centre of attention of researchers.

There are many proposals and specific models of scale symmetric Higgs scalar sector extended by scalar singlet \cite{Raidal2013, Raidal2014, Raidal2015, Raidal2016, Raidal2022, Shaposhnikov2008_1, Shaposhnikov2008_2, Shaposhnikov2011, Shaposhnikov2020, Shaposhnikov2021, Shaposhnikov2022,Ross2016_1, Ross2016_2, Ross2018, Ross2019}. Usually, such models are coupled to Einstein, or Brans-Dicke, gravity. What seems to be an attractive alternative solution is to use a natural extension of Riemannian geometry - the Weyl geometry \cite{Ghilencea2018_1, Ghilencea:2018_2, Ghilencea:2019, Ghilencea2019_WeylR^2inflation, Ghilencea2020_WeylvsPalatini, Ghilencea2021_SMinWeyl, Ghilencea2021_cosmological_evolution, Ghilencea:2022_non-metricity}, which avoids many problems encountered in Einstein formulation of gravity non-minimally coupled to scalar fields~\cite{Ghilencea:2022_non-metricity}. Using Weyl geometry, the new scalar sector, dilaton $\phi_0$, is not added \textit{ad hoc} to the Lagrangian, but has a geometric origin. Its vev generates all mass scales in the theory, so $M_P$ and $m_H$ arise naturally.

Scale symmetry does not have to be broken by dimensional regularization $\mu=const$ or dimensionful cutoff $\Lambda$. As indicated in previous proposals \cite{MSIGhilencea, Ghilencea2016_MSIsimilar, LOG_2loops, LOG, Ghilencea2017_2and3loops, Lalak_Olszewski_mu=fi0}, renormalization scale can be a function of dilaton $\mu = \mu(\phi_0)$. This way, the scale symmetry is preserved at the quantum level and after spontaneous scale symmetry breaking (SSSB), when dilaton acquires a vev $\langle\phi_0\rangle$, the subtraction scale $\mu(\langle\phi_0\rangle)$ is generated dynamically. Therefore, due to SSB another important mass scale is generated, not only $M_P$ and $m_H$. Preserving scale symmetry with $\mu=\mu(\phi_0)$ leads to new finite quantum corrections absent in standard Coleman-Weinberg results \cite{Coleman-Weinberg}. For a discussion of quantum corrections in this case, one should consult earlier work reported in \cite{MSIGhilencea, Ghilencea2016_MSIsimilar, LOG_2loops, LOG, Ghilencea2017_2and3loops, Lalak_Olszewski_mu=fi0}.

What has not been analysed yet are thermal corrections to classically scale invariant potential for Higgs scalar sector extended with dilaton. Such a study seems natural since if one wants to examine the cosmological evolution of the model, adding temperature corrections allows for the fact that the Universe is hot during expansion. Since the temperature $T$ is an explicit mass scale, it explicitly breaks the scale symmetry. Temperature corrections to the Higgs sector only, display electroweak symmetry restoration at high temperature~\cite{Sher}. Namely, Higgs expectation value is driven to zero and electroweak symmetry breaking (EWSB) occurs only after the system cools down. The shape of the thermally corrected potential for the Higgs suggests that EWSB is a second order, or weakly first order, phase-transition. The case may be different though if one adds additional new scalar singlet, dilaton. We analyse the results of time evolution of the fields in hot Universe to show the nature of EWSB in our model and to simulate the dynamical mass scale generation when dilaton settles at its final value $\langle\phi_0\rangle$.

The paper is organized as follows: Section \ref{Scale_sym_pot} provides analysis of classically scale symmetric potential used in the presented model and available parameter space. Section \ref{Therm_corr} includes a short description of finite temperature corrections and their application to considered theory. We show how temperature as a mass scale breaks scale symmetry explicitly and how it affects the potential changing the ground state. In Section \ref{Time_evolution} we present numerical simulations of time evolution of fields $\phi_0$ and $\phi_1$ in both zero temperature case and in hot Universe. We show that one can reasonably choose initial conditions in such a way that one obtains at late times of the evolution a realistic model fitting our present knowledge. In Appendix \ref{appendix} we describe basic properties of Weyl conformal geometry alone and coupled to matter fields. This gives additional motivation for the presence of the dilaton in our model, which can be considered as the low-energy Riemannian limit of Weyl conformal geometry.

\section{Scale symmetric potential}
\label{Scale_sym_pot}

Let us consider a scale symmetric Lagrangian for the Higgs neutral component $\phi_1$ and a new scalar singlet $\phi_0$, which we call dilaton. Coupling both fields with Einstein gravity via non-minimal couplings $\xi_i$ we have:
\begin{equation}
    \frac{\mathcal{L}}{\sqrt{g}} = -\frac{1}{12}\Big(\xi_0\phi_0^2+\xi_1\phi_1^2\Big)R+\frac{1}{2}\partial_{\mu}\phi_0\partial^{\mu}\phi_0+\frac{1}{2}\partial_{\mu}\phi_1\partial^{\mu}\phi_1-V(\phi_0,\phi_1),
    \label{Llowlimit}
\end{equation}
where $R$ is Ricci scalar in Riemannian geometry. In order to obtain scale symmetry at the tree level, one needs to assume a vanishing Higgs mass parameter $m_H^2=0$. In terms of $\phi_0$ and $\phi_1$ fields, scale symmetric potential at the classical level is of the form:
\begin{equation}
    V(\phi_0,\phi_1) = \lambda_0\phi_0^4+\lambda_1\phi_0^2\phi_1^2+\lambda_2\phi_1^4.
    \label{Vtreelevel}
\end{equation}
We choose certain hierarchy among couplings:
\begin{equation}
    \lambda_2\gg |\lambda_1| \gg \lambda_0
    \label{Lambdas_hierarchy}
\end{equation}
and $\lambda_2>0$, $\lambda_1<0$, $\lambda_0>0$, so the new dilaton sector is weakly coupled to the Higgs sector. The origin of the dilaton field can be found in Weyl conformal geometry, which at low energies comes down to Einstein gravity. All essential properties of such formulation are described in the Appendix \ref{appendix}. 

Let us consider FLRW metric $(1,-a(t)^2,-a(t)^2,-a(t)^2)$ with $\sqrt{g}=\sqrt{|\det g|}=a(t)^3$. Equations of motion from (\ref{Llowlimit}) for each field and $g_{\mu\nu}$ are:
\begin{equation}
    \begin{split}
        \phi_0: \quad & \quad \ddot{\phi_0}+3H\dot{\phi_0}+\frac{\xi_0}{6}\phi_0R+4\lambda_0\phi_0^3+2\lambda_1\phi_0\phi_1^2=0\\
        \phi_1: \quad & \quad \ddot{\phi_1}+3H\dot{\phi_1}+\frac{\xi_1}{6}\phi_1R+4\lambda_2\phi_1^3+2\lambda_1\phi_0^2\phi_1=0 \\
        g_{\mu\nu}: \quad & \quad \frac{1}{12}\Big(\xi_0\phi_0^2+\xi_1\phi_1^2\Big)R-\frac{1}{2}\dot{\phi_0}^2-\frac{1}{2}\dot{\phi_1}^2+2\big(\lambda_0\phi_0^4+\lambda_1\phi_0^2\phi_1^2+\lambda_2\phi_1^4\big)=0\\
    \end{split}
    \label{EOMS}
\end{equation}
where $H=\frac{\dot{a}}{a}$ is Hubble parameter. Stationary solutions give a flat direction:
\begin{equation}
    \langle\phi_1^2\rangle = -\frac{\lambda_1}{2\lambda_2}\langle\phi_0^2\rangle, \qquad \lambda_0=\frac{\lambda_1^2}{4\lambda_2}, \qquad \langle R\rangle = 0,
\end{equation}
where the $\lambda_0$ dependence comes from the condition of zero cosmological constant at the ground state:
$$
V(\langle\phi_0\rangle,\langle\phi_1\rangle)=0.
$$
Since the theory is scale symmetric, only ratios of mass scales can be determined and $\langle\phi_0\rangle$ is arbitrary. With (\ref{Lambdas_hierarchy}) we have the hierarchy $\langle\phi_0\rangle\gg\langle\phi_1\rangle$. When $\phi_0$ acquires its vev, scale symmetry is broken and flat direction no longer exists. Because $\langle\phi_1\rangle$ is proportional to $\langle\phi_0\rangle$, dilaton generates Higgs vev and mass, so it can be considered as origin of mass scales.

The mass matrix:
\begin{equation}
M^2 = 
\left( \begin{array}{cc}
\lambda _1 \left(2 \phi_1^2+\frac{3 \lambda _1 }{\lambda _2}\phi_0^2\right) & 4\lambda_1\phi_1\phi_0 \\
4\lambda_1\phi_1\phi_0 & 2 \left(6  \lambda _2\phi_1^2+\lambda _1 \phi_0^2\right)
\end{array}
\right),
\label{M2}
\end{equation}
has two eigenvalues (at the ground state):
\begin{equation}
m_G^2 = 0, \qquad m_H^2 = -4\lambda_1\Big(1-\frac{\lambda_1}{2\lambda_2}\Big)\langle\phi_0^2\rangle,
\end{equation}
so one of the eigenstates is massless (Goldstone associated with scale symmetry and flat direction).

\subsection{Higgs potential parameters and Planck mass:}
\label{Higgspotpar}

One can determine what values of $\lambda_1$, $\lambda_2$, $\xi_0$ and $\xi_1$ are available in our theory. First, we want the hierarchy (\ref{Lambdas_hierarchy}) and two conditions to be fulfilled:
\begin{equation}
m_H^2 = (125\textrm{ GeV})^2, \qquad \langle \phi_1 \rangle = 250 \textrm{ GeV}
\end{equation}
and in our model we have:
\begin{equation}
    m_H^2 = -4\lambda_1\Big(1-\frac{\lambda_1}{2\lambda_2}\Big)\langle\phi_0^2\rangle, \quad \langle\phi_1^2\rangle = -\frac{\lambda_1}{2\lambda_2}\langle\phi_0^2\rangle.
\end{equation}
Satisfying all the conditions we get:
\begin{equation}
\lambda_2 = \frac{1}{32}\Big(1+16\lambda_1\Big), \qquad -\frac{1}{48}\leq \lambda_1 \leq 0
\label{lambda2and1}
\end{equation}
and example values are:
\begin{equation}
    \lambda_2(\lambda_1= -10^{-6})\approx \lambda_2(\lambda_1=-10^{-11})\approx 0.03125.
\end{equation}
Required value of $\langle\phi_0\rangle$ is then:
\begin{equation}
    \langle\phi_0^2\rangle = -\frac{2\lambda_2}{\lambda_1}\langle\phi_1^2\rangle = -\frac{2\lambda_2}{\lambda_1}\cdot (250 \textrm{ GeV})^2.
    \label{dilaton_ground}
\end{equation}

Then, we want the Planck mass scale to be generated by $\phi_i$ fields, as in (\ref{MPlanck}):
\begin{equation}
    \frac{1}{6}\Bigg(\xi_0-\frac{\lambda_1}{2\lambda_2}\xi_1\Bigg)\langle\phi_0^2\rangle = M_{Planck}^2.
    \label{Mpl_ground}
\end{equation}
This will force constraints on values of $\xi_i$ couplings. In realistic models, \cite{Ghilencea:2018_2}, we should have $\xi_1\ll\xi_0$. Using (\ref{lambda2and1}), (\ref{dilaton_ground}) and (\ref{Mpl_ground}) we obtain relation of couplings:
\begin{equation}
    \lambda_1 = \frac{-0.0625\cdot\xi_0}{\xi_0-\xi_1+1.43\cdot10^{34}}.
\end{equation}
Example values:
\begin{equation}
    \begin{split}
        \xi_0&=10^{5},\quad \xi_1=0.1 \quad \Rightarrow \quad \lambda_1 = -4.37\cdot10^{-31}, \\
        \xi_0&=10^{10},\quad \xi_1=0.1 \quad \Rightarrow \quad \lambda_1 = -4.37\cdot10^{-26}, \\
        \xi_0&=10^{15},\quad \xi_1=0.1 \quad \Rightarrow \quad \lambda_1 = -4.37\cdot10^{-21}.
    \end{split}
\end{equation}

\section{Temperature corrections}
\label{Therm_corr}

\subsection{Quantum Field Theory at finite temperature:}

To obtain temperature corrections, one adds to potential temperature dependent parts \cite{Carrington, quiros, QFTaFT}:
\begin{equation}
   V(\phi_0,\phi_1)\rightarrow V(\phi_0,\phi_1) + \delta V_T(\phi_0,\phi_1,T) +\delta V_{ring}(\phi_0,\phi_1,T).
\end{equation}
The $\delta V_T$ stands for standard temperature corrections of first order:
\begin{equation}
    \delta V_T(\phi_0,\phi_1,T)   = \frac{T^4}{2\pi^2}\Bigg[\sum_{i=\textrm{bosons}}n_i\cdot J_B\Big(\frac{m_i^2(\phi_k)}{T^2}\Big) +\sum_{j=\textrm{fermions}}n_j\cdot J_F\Big(\frac{m_j^2(\phi_k)}{T^2}\Big)\Bigg],
    \label{dVT}
\end{equation}
where $n_i$ and $n_j$ are numbers of degrees of freedom of considered boson or fermion particle with field-dependent mass $m_i(\phi_k)$ and $J_B$ and $J_F$ are thermal bosonic (B) or fermionic (F) functions defined as follows:
\begin{equation}
    J_B\Big(\frac{m^2}{T^2}\Big) = \int_0^{\infty} dx\cdot x^2\log\Big(1-e^{-\sqrt{x^2+\frac{m^2}{T^2}}}\Big),
\end{equation}
\begin{equation}
    J_F\Big(\frac{m^2}{T^2}\Big) = \int_0^{\infty} dx\cdot x^2\log\Big(1+e^{-\sqrt{x^2+\frac{m^2}{T^2}}}\Big).
\end{equation}
One should also add infrared contributions from higher order diagrams \cite{Carrington}, which are of the same order as corrections from (\ref{dVT}):
\begin{equation}
    \delta V_{ring} = -\frac{T}{12\pi}\Big(m_{eff}(\phi_i,T)^3-m_i(\phi_i)^3\Big),
\end{equation}
which is the so-called ring improvement of the potential and is sufficient for high temperatures, when $(m/T)\ll~1$. Temperature-dependent masses $m_{eff}(\phi_i,T)$, can be obtained from high temperature expansion of:
\begin{equation}
    V+\delta V_T\Big|_{m/T\ll1}.
    \label{hTe}
\end{equation}
 
 For the case of the present paper we assume, following \cite{Postma1,Postma2}, that the one-loop equivalence holds for the effective potential in the Einstein frame and in the Jordan frame and use for simplicity the Jordan frame.  As for the thermal corrections, we compute them in the Jordan frame, where the matter Lagrangian is renormalizable.
 
\subsubsection*{Particle content and thermal masses}

Obviously, the degrees of freedom $\phi_0$ and $\phi_1$ are present in our theory, and the mass eigenstates are their mixture. So we have two neutral scalars $G$ (massless Goldstone) and $H$ (massive "Higgs") with $n_G=n_H=1$. The masses of this sector are the eigenvalues of~(\ref{M2}):
\vspace{-10pt}
\begin{equation}
    \begin{split}
        m_G^2 = & 2  \lambda _1\phi_1^2+\mathcal{O}(\lambda_1^2) \\
        m_H^2 = & 12 \lambda _2\phi_1^2+ 2\lambda _1 \phi_0^2+\mathcal{O}(\lambda_1^2).
    \end{split}
\end{equation}
However, there are particles in SM, which give important contributions to (\ref{dVT}). These are:
\begin{itemize}
    \item[-] $W^{\pm}$ boson: $m_W^2 = \frac{1}{4}g_2^2\phi_1^2$,  $n_w = 6$,
    \item[-] $Z$ boson: $m_Z^2 = \frac{1}{4}(g_1^2+g_2^2)\phi_1^2$,  $n_Z=3$
    \item[-] top quark: $m_t^2 = \frac{1}{2}h_t^2\phi_1^2$,  $n_t = -12$,
\end{itemize}
where $g_1\approx0.35$, $g_2\approx0.65$ and $h_t\approx1$ are correspondingly weak, strong and top yukawa coupling 
constants\footnote{It is well known that the Higgs effective potential in the Standard Model, calculated perturbatively, generically suffers from infrared (IR) divergences when the field-dependent tree-level mass of the Goldstone bosons in the Higgs doublet goes vanishing. Here we follow the analysis given in \cite{Martin} and \cite{Elias-Miro} and assume that such divergences can be cured by a resummation of  IR-problematic terms to any order and neglect these troublesome contributions.}.

From (\ref{hTe}) we obtain the mass matrix:
\begin{equation}
\begin{split}
    \left(
    \begin{array}{cc}
        m_{00} & m_{10}  \\
        m_{01} & m_{11}
    \end{array}
\right)_{eff}
= &
\left(
    \begin{array}{cc}
        \frac{\partial^2 V}{\partial\phi_0^2} & \frac{\partial^2 V}{\partial\phi_1\partial\phi_0}  \\
        \frac{\partial^2 V}{\partial\phi_0\partial\phi_1} &  \frac{\partial^2 V}{\partial\phi_1^2}
    \end{array}
\right)
+ \\
 & + \left(
\begin{array}{cc}
 \Big(\frac{\lambda_1}{6}+\frac{\lambda_1^2}{4\lambda_2}\Big)T^2 & 0 \\
 0 & \Big(\lambda_2+\frac{\lambda_1}{6}+\frac{g_1^2}{16}+\frac{3g_2^2}{16}+\frac{h_t^2}{4}\Big)T^2 \\
\end{array}
\right).
\end{split}
\label{M2thermal}
\end{equation}
The thermal masses $m_{eff}^2$ for scalars are the eigenvalues of this matrix:
\begin{equation}
    \begin{split}
        \big(m_G^2\big)_{eff} = & 2\lambda_1\phi_1^2+\frac{\lambda_1}{6}T^2+\mathcal{O}(\lambda_1^2), \\
        \big(m_H^2\big)_{eff} = & 12\lambda_2\phi_1^2+2\lambda_1\phi_0^2+\Big(\lambda_2+\frac{\lambda_1}{6}+\frac{g_1^2}{16}+\frac{3g_2^2}{16}+\frac{h_t^2}{4}\Big)T^2+\mathcal{O}(\lambda_1^2).
    \end{split}
\end{equation}
It should be noted, that the dependence of the temperature corrections on $\phi_0$ enters via the tree-level mass terms in the scalar sector. Moreover, since the hierarchy of scales demands the hierarchy of couplings, one finds in the present case $\lambda_0 =  \frac{\lambda_1^2}{4 \lambda_2} \ll \vert \lambda_1 \vert  \ll \lambda_2 $ and the dependence on $\phi_0$ starts at the linear order in a small coupling $\lambda_1$. At this point, we refrain from discussing the issue of thermal equilibrium of the whole system and concentrate on the analysis of the thermal effective potential.

\subsection{Symmetry breaking at high temperature}
The scale symmetry breaking can be discussed reliably at the leading level of high temperature expansion, where 
\begin{equation}
    V_{eff} = V_{T=0}+\frac{1}{2}\phi_1^2\cdot\Big(\lambda_2+\frac{\lambda_1}{6}+\frac{g_1^2}{16}+\frac{3g_2^2}{16}+\frac{h_t^2}{4}\Big)T^2+\frac{1}{2}\phi_0^2\cdot\frac{\lambda_1}{6}T^2 = V_{T=0}+\frac{\gamma T^2}{2} \phi_1^2 + \frac{\lambda_1 T^2}{12} \phi_0^2.
\end{equation}
One should note that the coefficients of the two terms quadratic in the temperature are completely independent, as in the case of $\phi_1$, which plays the role of the Higgs field, the thermal corrections are dominated by gauge couplings and by the coupling to the top quark, whereas in the case of the $\phi_0$ they are proportional to the coupling of the mixing term in the scalar sector. As the result, the proportionality of the two scalar equations of motion which holds at $T=0$ gets broken by the term: 
\begin{equation}
    \Big( \frac{\lambda_1}{6} - \gamma \Big) T^2 \neq 0. 
\end{equation}
This is the amount of the scale symmetry breaking by the finite temperature effects. As the result, the only consistent solution to the corrected equations of motion becomes at this order
\begin{equation}
    \phi_1 = 0, \qquad \phi_0^2 = - \frac{\lambda_2}{6 \lambda_1} T^2 .
\end{equation}
This shows, that at finite temperature the minimum of the potential picks up a finite expectation value of the dilaton underlying the fact that the scale symmetry remains broken, and the scale of the breaking given by the vev of the dilaton is proportional to the temperature - the new scale in the system. However, this indicates, that when the temperature goes to zero, the symmetry gets restored and the system goes into the unbroken phase, since the vevs of both scalars seem to be led to the origin. 
One should note that this would restore also the electroweak symmetry, which requires a nonvanishing vev of $\phi_0$. The point is that it is this vev which multiplied by the negative coupling $\lambda_1$ plays the role of the negative mass squared term in the Higgs sector. This would suggest a symmetric, unrealistic, vacuum emerging from the hot phase of the universe. 
However, the situation is more subtle. The point is that the Higgs field $\phi_1$ easily comes to equilibrium with the rest of the universe through interaction with the Standard Model matter and gauge fields, which despite the fact that its thermal average seems to vanish, produces a large rms 
value of the order of:
\begin{equation}
    \langle\phi_1^2 \rangle_{T, p} = T \frac{p^3}{\omega^2_p}
    \label{Tpw}
\end{equation}
per decade. For high temperatures and small masses this can be approximated as $T^2$, and produces a large repulsive force in the equation of motion of $\phi_0$ due to the mixing term in the potential:
\begin{equation}
 \delta_m V = \lambda_1 \phi_0^2 \phi_1^2 \rightarrow  \lambda_1 T^2 \phi_0^2,  
\end{equation}
giving in the EOM the contribution:
\begin{equation}
 - \frac{ \partial \delta_m V}{ \partial \phi_0} = - 2 \lambda_1  T^2 \phi_0,  
\end{equation}
which drives the dilaton away from the origin. In addition, as discussed later, the dynamical thermal equilibrium in the scalar sector, perhaps after the point of quasi-thermal initial production, is not to be maintained at the later stages of the evolution of the universe. Hence, in the realistic physical system the origin will not be achieved globally and there will be in the universe domains characterized by large expectation value of the dilaton, and hence the Higgs, which when the temperature drops will have a chance to evolve dynamically into the zero temperature vacuum with spontaneously broken scale symmetry and electroweak symmetry. 
The possible late time dynamics of such systems shall be discussed in the Section \ref{Time_evolution}.

A remark is in order here. We have assumed above the manifestly scale-invariant regularization as described in \cite{MSIGhilencea, Ghilencea2016_MSIsimilar, LOG_2loops, LOG, Ghilencea2017_2and3loops, Lalak_Olszewski_mu=fi0}. Then, due to scale invariance at the level of quantum corrections, the one-loop potential can be approximated by the tree-level formula with couplings understood as running couplings. However, in the present case, where the hierarchy is based on the ratio of very small couplings, the non-thermal perturbative quantum corrections are small as proportional to higher powers of small couplings with respect to temperature corrections. Possible additional perturbative contributions violating explicitly scale symmetry, other than temperature effects, will shift the position of the scalar vevs, but note that the thermal shift of the $\phi_0^2$ is proportional to a very large ratio of couplings $\frac{\lambda_2}{|\lambda_1|} \gg 1$, hence other perturbative shifts would be typically subdominant unless the temperature is very low. 
Here we concentrate on the role of temperature corrections, hence we assume the scale invariance at the loop level. 

\subsubsection{Numerical analysis}

Here we show numerical results of how thermal corrections change the potential. The results nicely illustrate the approximate analytic picture presented in the previous section. To make plots more clear, we choose the value of the coupling to be $\lambda_1=-10^{-6}$.

\begin{figure}[!h]
\vspace{10pt}
     \centering
     \begin{subfigure}[b]{0.48\textwidth}
         \centering
         \includegraphics[width=\textwidth]{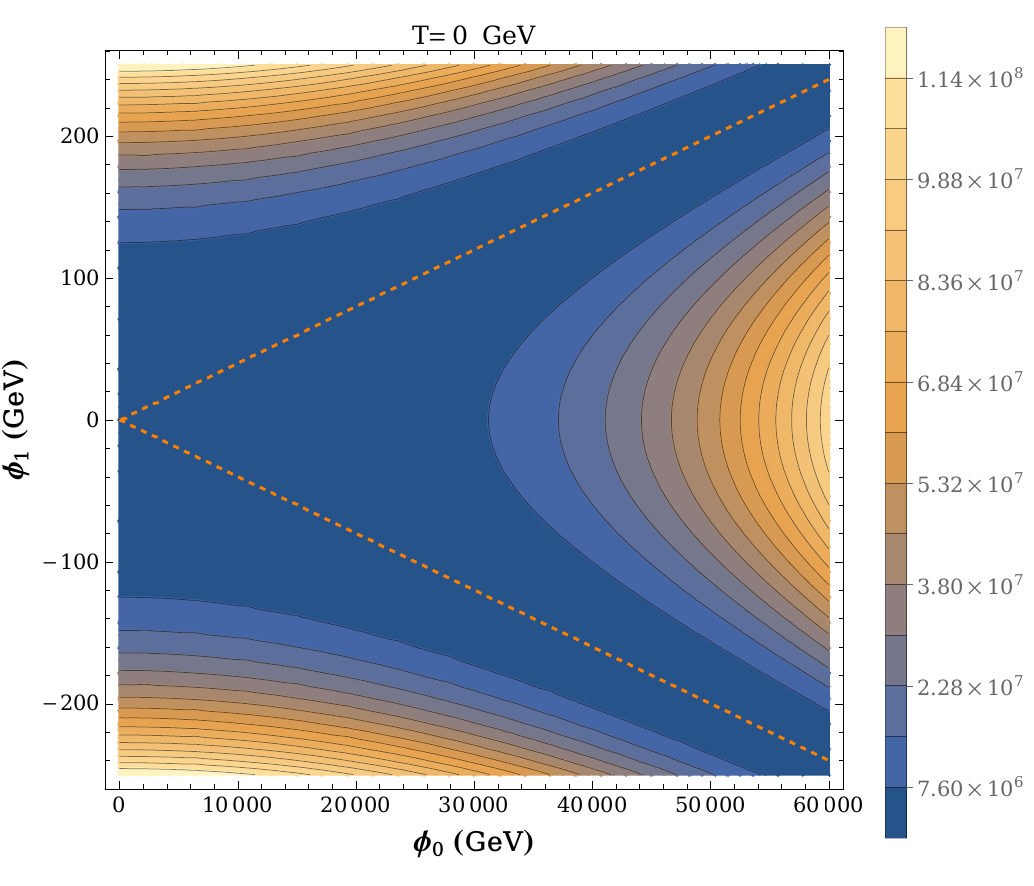}
     \end{subfigure}
     \hfill
     \begin{subfigure}[b]{0.48\textwidth}
         \centering
         \includegraphics[width=\textwidth]{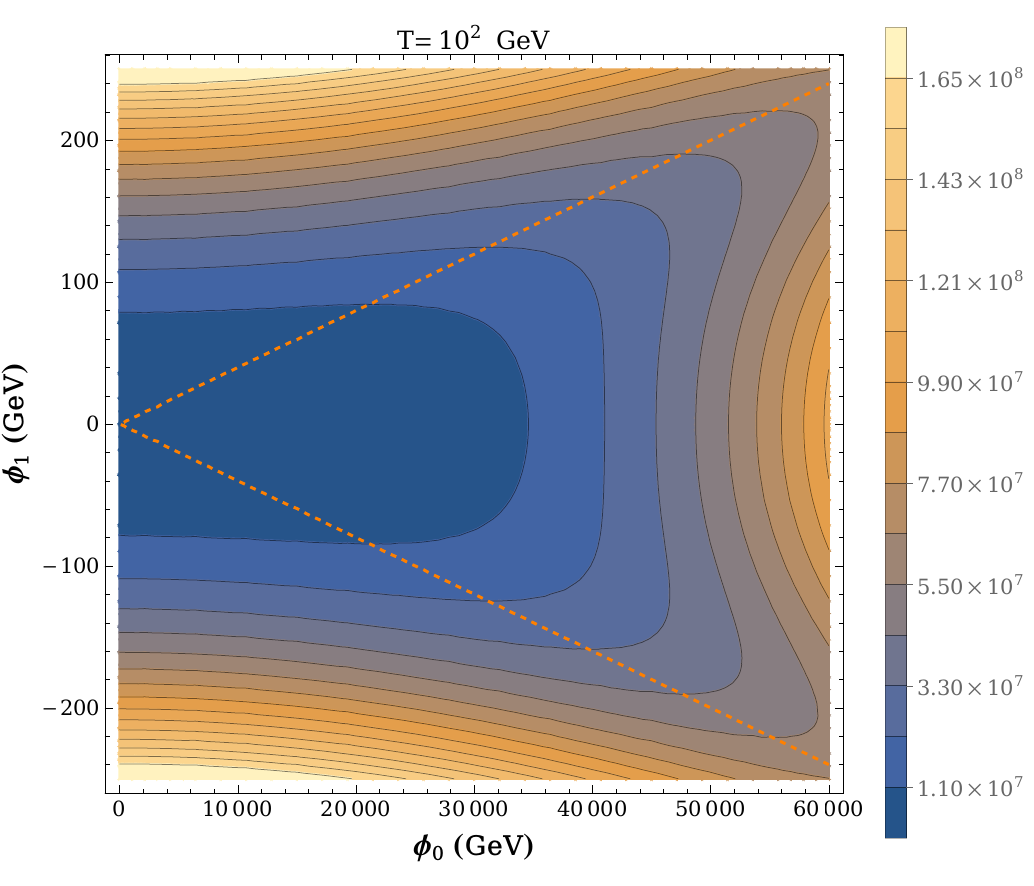}
    \end{subfigure}
    \\
    \begin{subfigure}[b]{0.48\textwidth}
         \centering
         \includegraphics[width=\textwidth]{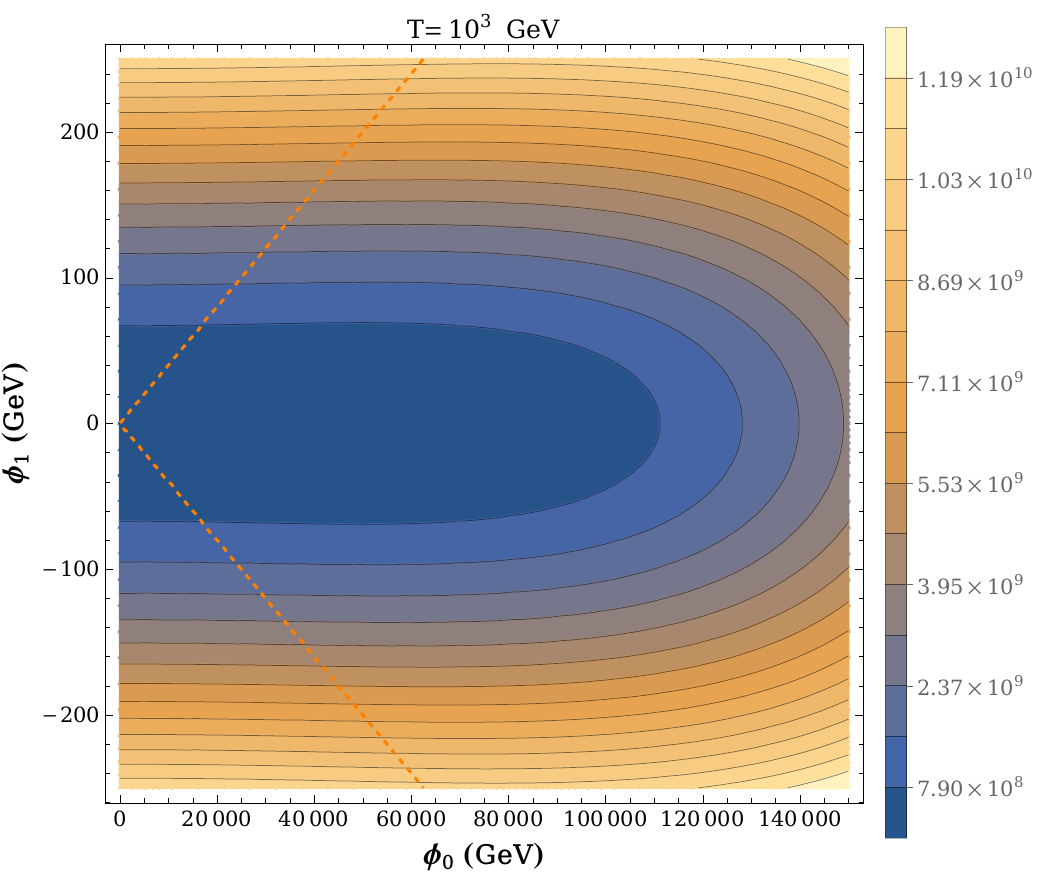}
     \end{subfigure}
     \hfill
     \begin{subfigure}[b]{0.48\textwidth}
         \centering
         \includegraphics[width=\textwidth]{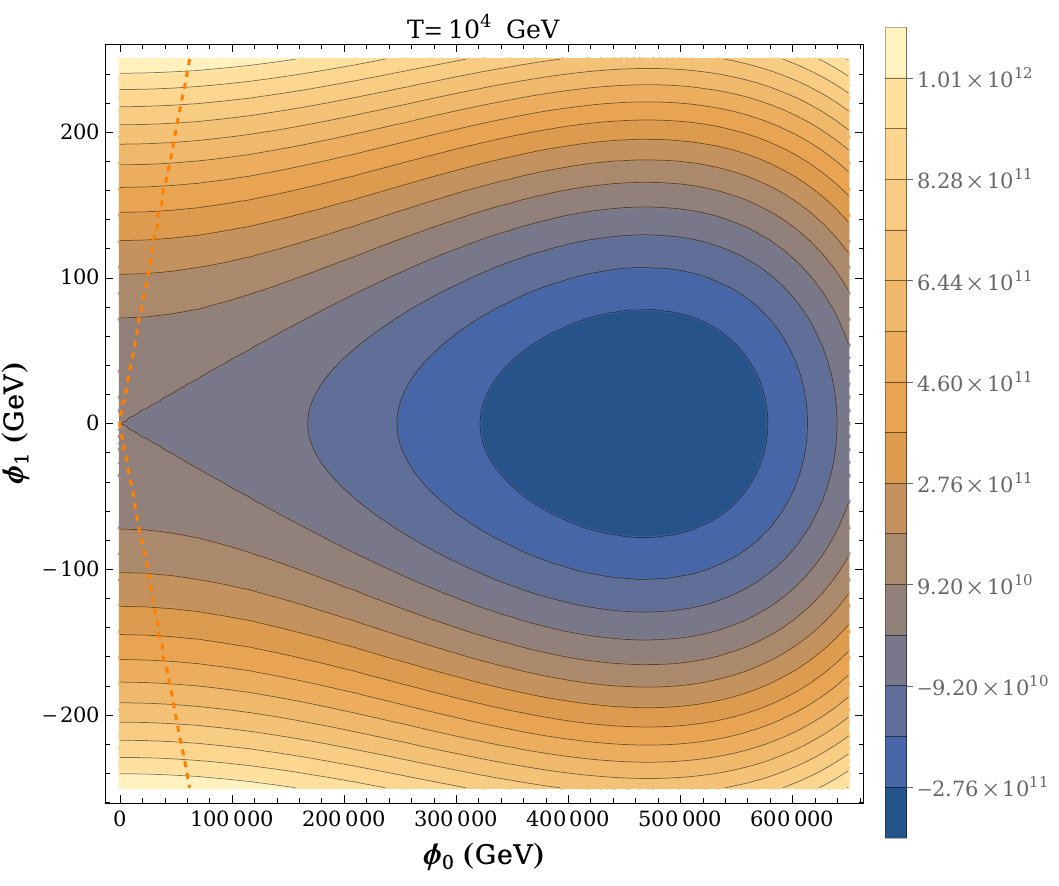}
    \end{subfigure}
    \caption{Plots of $V_{full}(\phi_0,\phi_1,T)$ for different temperatures and $\lambda_1 = -10^{-6}$. Orange dashed line marks flat direction $\phi_1^2 = -\frac{\lambda_1}{2\lambda_2}\phi_0^2$. It is easy to see that as the temperature increase, the flat direction no longer exists and the scale symmetry is broken.}
    \label{Vfullplots}
    \vspace{10pt}
\end{figure}

\begin{figure}[!t]
     \centering
     \begin{subfigure}[b]{0.48\textwidth}
         \centering
\includegraphics[width=\textwidth]{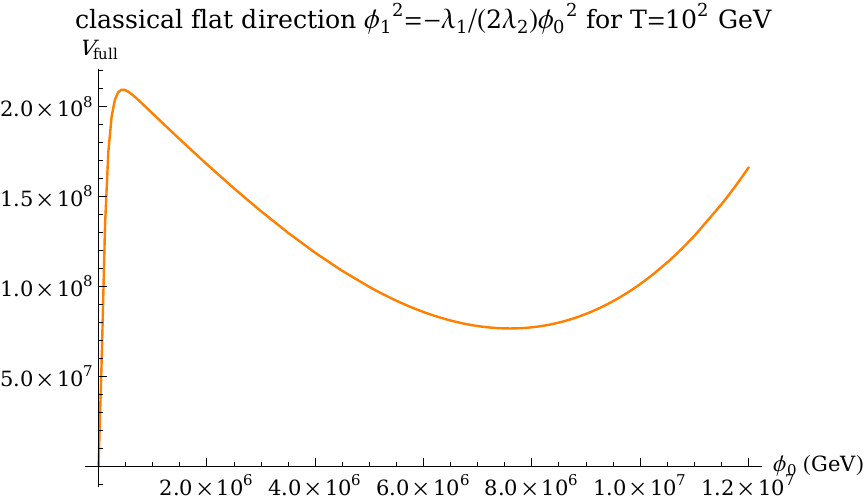}
     \end{subfigure}
     \begin{subfigure}[b]{0.48\textwidth}
         \centering
\includegraphics[width=\textwidth]{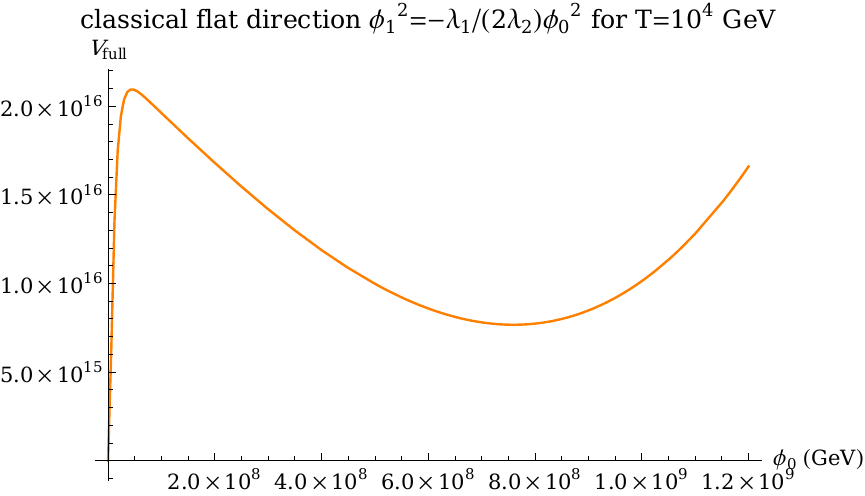}
    \end{subfigure}
    \caption{Classical flat direction is spoiled for non-zero temperatures.}
\label{flatdir}
\vspace{15pt}
\end{figure}

\begin{figure}[!t]
     \centering
     \begin{subfigure}[b]{0.48\textwidth}
         \centering
         \includegraphics[width=\textwidth]{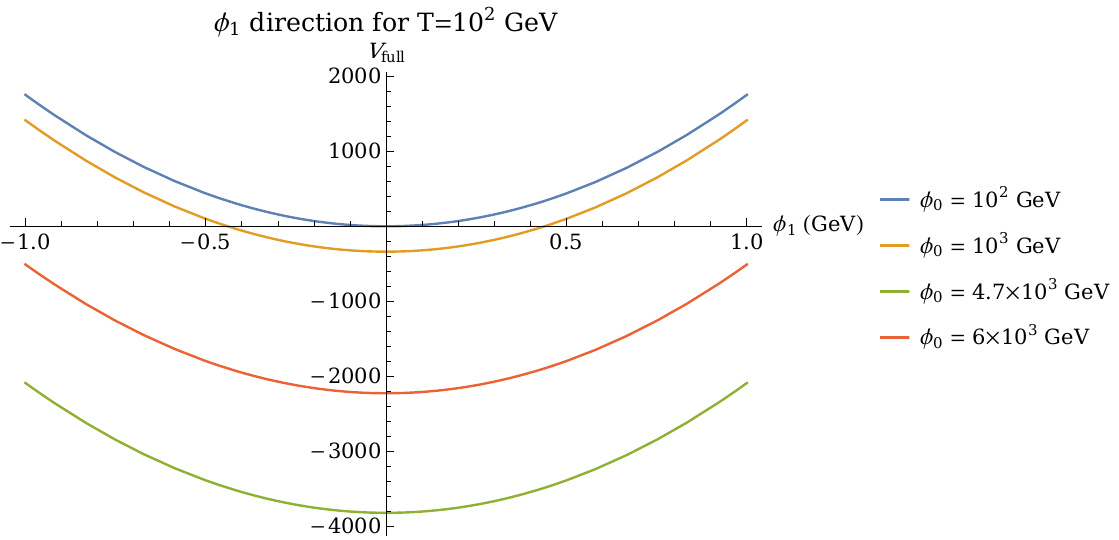}
     \end{subfigure}
     \begin{subfigure}[b]{0.48\textwidth}
         \centering
         \includegraphics[width=\textwidth]{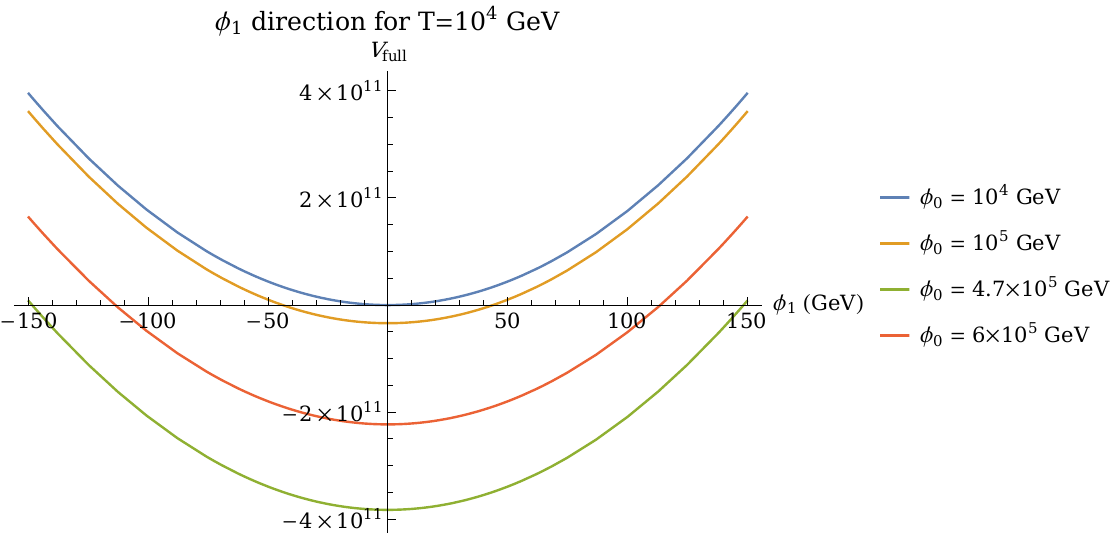}
    \end{subfigure}
    \caption{$\phi_1$ direction of $V_{full}(\phi_0,\phi_1,T)$ for 2 different temperatures and various $\phi_0$ values. The lowest curve corresponds to $\phi_0$ from high temperature minimum (\ref{HTm}).}
\label{f1dir}
\end{figure}

\begin{figure}[!t]
     \centering
     \begin{subfigure}[b]{0.48\textwidth}
         \centering
         \includegraphics[width=\textwidth]{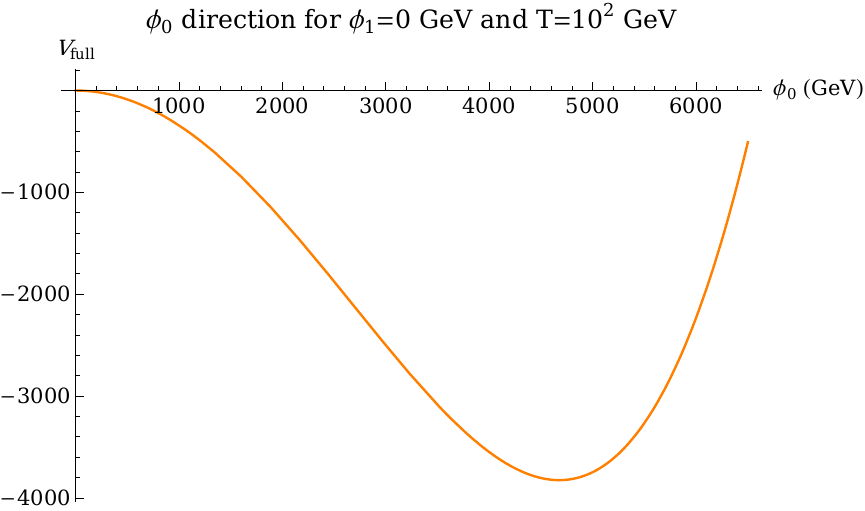}
     \end{subfigure}
     \begin{subfigure}[b]{0.48\textwidth}
         \centering
         \includegraphics[width=\textwidth]{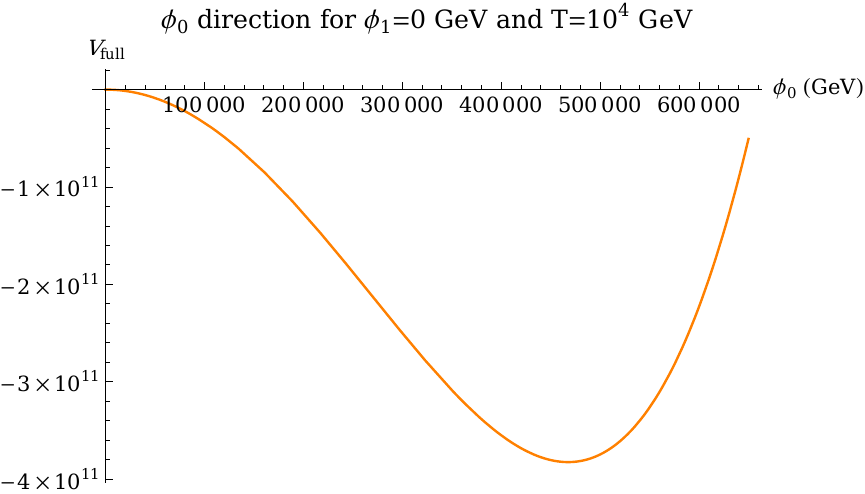}
    \end{subfigure}
    \caption{$\phi_0$ direction of $V_{full}(\phi_0,\phi_1,T)$ for 2 different temperatures and $\phi_1=0$. There's a visible minimum at $\phi_0\approx 4.7\cdot 10\cdot T$, which corresponds to equation (\ref{HTm}).}
\label{f0dir}
\end{figure}

As one expects, temperature corrections break the scale symmetry. In Figure (\ref{Vfullplots}) one can see that for higher temperatures, the flat direction no longer exists (Figure \ref{flatdir}) and there is visible a local minimum. We made it better visible in Figures \ref{f1dir} and \ref{f0dir}, where $\phi_i$ directions are plotted. Temperature corrections drive $\phi_1$ to zero value. The dilaton field minimum then becomes:
\begin{equation}
\begin{split}
    \phi_0^2 = & \Bigg[\frac{\big(9.89\alpha-3.63\lambda_2-6.91\cdot10^{-16}g_1^2-2.76\cdot10^{-15}h_t^2\big)}{\big(-39.48\alpha+\lambda_2(\alpha\log(-\lambda_1)+3.48\alpha+43.53)\big)}+\\
    &+\frac{\lambda_2\big(6.58\alpha-43.53\lambda_2-2.72g_1^2-8.16g_2^2-10.88h_t^2\big)}{\lambda_1\big(-39.48\alpha+\lambda_2(\alpha\log(-\lambda_1)+3.48\alpha+43.53)\big)}\Bigg]\cdot T^2,
    \end{split}
    \label{HTm}
\end{equation}
where
\begin{equation}
    \alpha = \sqrt{48\lambda_2+3g_1^2+9g_2^2+12h_t^2}.
\end{equation}

\subsection{Thermal equilibrium}
\label{therm_eq}
To discuss physical applications of the scale symmetric models, one needs to consider the issue of thermal equilibrium. 

First, let us investigate in which temperature range the $\phi_0$ is in thermal equilibrium. To do so, one needs to calculate a cross-section \cite{kinematics}:
\begin{equation}
\frac{d\sigma}{d t} = \frac{1}{64\pi\cdot s}\frac{1}{|\vec{p}_{1CM}|^2}\big|\mathcal{M}\big|^2,
\end{equation}
where $s$ and $t$ are Mandelstam variables and $\vec{p}_{1CM}$ is momentum of an in-going particle in the centre of mass frame. All essential steps and formulae are to be found in \cite{kinematics}. We also use the following approximations:
\begin{equation}
    m_{\phi_0}\approx m_G=0, \qquad m_{\phi_1}\approx m_H=125\textrm{ GeV}.
\end{equation}
The only $\phi_0\phi_1$ interaction present in $V$ is $\lambda_1\phi_0^2\phi_1^2$ with $|\mathcal{M}|^2 = 16\lambda_1^2$. So there are two processes:
\begin{itemize}
\item[•] INELASTIC $\phi_0\phi_1\rightarrow\phi_0\phi_1$:
$$
\sigma(s) = \frac{\lambda_1^2}{\pi\cdot s},
$$
\item[•] ELASTIC $\phi_1\phi_1\rightarrow\phi_0\phi_0$:
$$
\sigma(s) = \frac{\lambda _1^2}{\pi  \sqrt{s \left(s-4 m_H^2\right)}}.
$$
\end{itemize}

\begin{figure}[!t]
     \centering
     \begin{subfigure}[b]{0.51\textwidth}
         \centering
        \includegraphics[width=\textwidth]{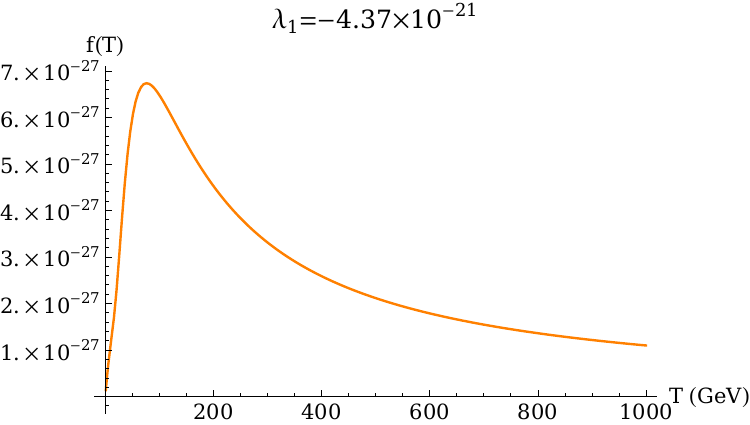}
     \end{subfigure}
     \begin{subfigure}[b]{0.48\textwidth}
         \centering
         \includegraphics[width=\textwidth]{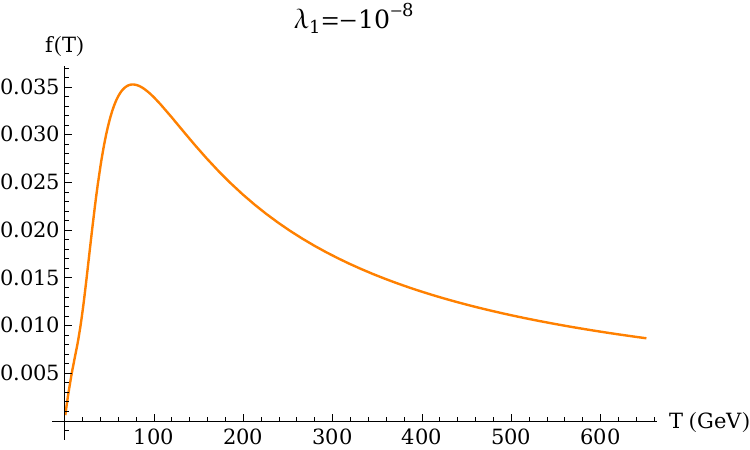}
    \end{subfigure}
    \\
    \begin{subfigure}[b]{0.48\textwidth}
         \centering
         \includegraphics[width=\textwidth]{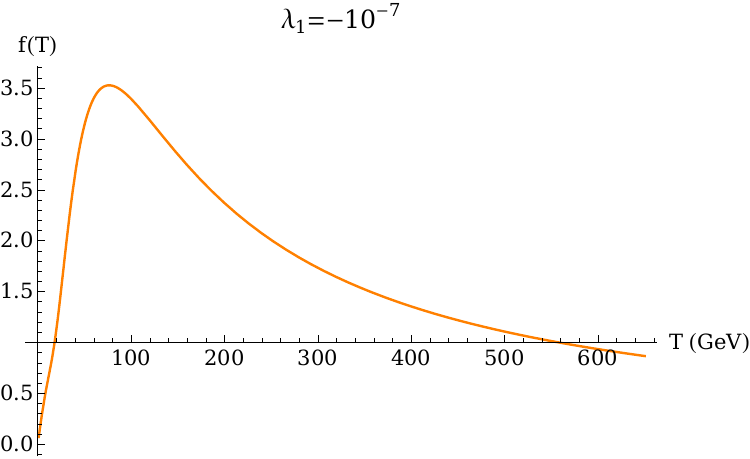}
     \end{subfigure}
     \begin{subfigure}[b]{0.48\textwidth}
         \centering
         \includegraphics[width=\textwidth]{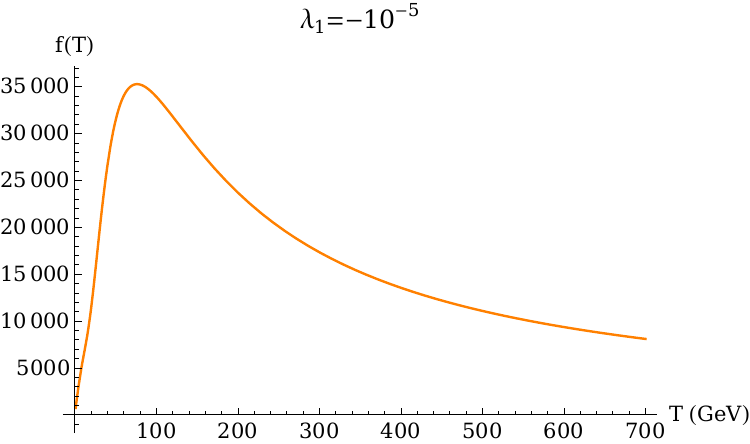}
     \end{subfigure}
    \caption{Ratio $f(T) =(T^3\langle\sigma v\rangle)/H $ as a function of temperature for different $\lambda_1$. $\phi_0$ field can reach thermal equilibrium for sufficiently large $|\lambda_1|$ value. }
    \label{GoverH}
\end{figure}

Next, thermally averaged cross-section for elastic process $\phi_0\phi_1\rightarrow\phi_0\phi_1$ is of the form~\cite{TACS}:
\begin{equation}
\langle\sigma\cdot v\rangle_{EL} = \frac{1}{16 K_2(m_H/T)\cdot m_H^2 T^3}\int_{m_H^2}^{\infty}ds\cdot \lambda(s,m_H,0)\frac{\sigma(s)}{\sqrt{s}}K_1(\sqrt{s}/T)
\end{equation}
and for inelastic process $\phi_1\phi_1\rightarrow\phi_0\phi_0$ \cite{TACS}:
\begin{equation}
\langle\sigma\cdot v\rangle_{INEL} = \frac{1}{8 K_2(m_H/T)^2\cdot m_H^4 T}\int_{m_H^2}^{\infty}ds\cdot \lambda(s,m_H,m_H)\frac{\sigma(s)}{\sqrt{s}}K_1(\sqrt{s}/T),
\end{equation}
where $K_i(x)$ is a modified Bessel function of $i-th$ kind and
$$
\lambda(x,y,z) = \big[x-(y+z)^2\big]\cdot \big[x-(y-z)^2\big].
$$

Particles are in thermal equilibrium as long as their interaction rate $\Gamma$ satisfies:
\begin{equation}
\Gamma \gtrsim H,
\end{equation}
where $H$ is Hubble parameter, and for radiation dominated era it is of the form
\begin{equation}
H(T) = \frac{1}{\sqrt{3}m_{Pl}}\sqrt{\frac{\pi^2}{30}g_{\ast}T^4},
\label{H(T)dep}
\end{equation}
where $g_{\ast}$ is the number of effective degrees of freedom of relativistic particles in hot Universe and Planck mass is $m_{Pl}=2.4354\cdot 10^{18}$ GeV. Interaction rate is proportional to the thermally averaged cross-section $\Gamma  \sim T^3 \langle\sigma v\rangle$, hence we can examine ratio $ (T^3\langle\sigma v\rangle)/H = f(T)$ as a function of temperature, where
\begin{equation}
    \langle\sigma v\rangle = \langle\sigma v\rangle_{EL} + \langle\sigma v\rangle_{INEL}.
\end{equation}
For temperatures above $0.1$ GeV we have $g_{\ast}\approx 100$ \cite{g_astar}. We show the results for different $\lambda_1$ on Figure~\ref{GoverH}. Dilaton $\phi_0$ can reach the equilibrium state for $\lambda_1\gtrsim-10^{-7}$.

\section{Time evolution of the fields}
\label{Time_evolution}

In this section, we shall investigate numerically the evolution of the scale symmetric scalar sector in the expanding universe. We show that there exist reasonable initial conditions which lead to the physically relevant vacuum configuration at the very late stages of the evolution. 

\subsection{Zero temperature}

We use equations of motion (\ref{EOMS}) with relation $R=12H^2+6\dot{H}$:
\begin{equation}
    \begin{split}
    &\ddot{\phi_0}+3H\dot{\phi_0}+2\xi_0\phi_0^2H^2+\xi_0\phi_0^2\dot{H}+4\lambda_0\phi_0^3+2\lambda_1\phi_0\phi_1^2=0\\
        &\ddot{\phi_1}+3H\dot{\phi_1}+2\xi_1\phi_1^2H^2+\xi_1\phi_1^2\dot{H}+4\lambda_2\phi_1^3+2\lambda_1\phi_0^2\phi_1=0 \\
        &\frac{1}{2}\Big(\xi_0\phi_0^2+\xi_1\phi_1^2\Big)\big(2H+\dot{H}\big)-\frac{1}{2}\dot{\phi_0}^2-\frac{1}{2}\dot{\phi_1}^2+2\big(\lambda_0\phi_0^4+\lambda_1\phi_0^2\phi_1^2+\lambda_2\phi_1^4\big)=0\\
    \end{split}
    \label{EOMSagain}
\end{equation}
We treat the Hubble parameter as an independent variable, which dynamics is ruled by above equations. Fixed points of (\ref{EOMSagain}) are:
\begin{equation}
    \langle\phi_1^2\rangle = -\frac{\lambda_1}{2\lambda_2}\langle\phi_0^2\rangle, \qquad \langle H\rangle =0.
\end{equation}

\subsubsection*{Example evolution solutions}

All initial conditions used in simulations are provided under plots in the figure captions. Two initial Hubble parameter $H(0)=H_0$ values were chosen: 0.1 GeV and 0.5 GeV. Since $H$ is a magnitude of expansion, hence it can be interpreted as a parameter describing how fast $\phi_i$ fields lose their energy. Bigger $H_0$ values result in faster velocity $\dot{\phi}_i$ damping and lower final values $\langle\phi_i\rangle$. We choose two parameter spaces: one with realistic $\lambda_i$ and $\xi_i$ values, fullfiling requirements from section \ref{Higgspotpar} and one with bigger $|\lambda_1|$ and smaller $\xi_0$. In both cases, $\phi_i$ fields stop to evolve after sufficiently long time and settle at some point of the flat direction $\langle\phi_1^2\rangle=-\frac{\lambda_1}{2\lambda_2}\langle\phi_0^2\rangle$. Of course, the arrow of the initial velocity might play a role, but this is a direct choice and we always pick up the velocity direction, which enhances the effects.

\newpage

\subsubsection*{Realistic model:} $\lambda_2=0.03125$, $\lambda_1=-4.37\cdot10^{-26}$, $\xi_0=10^{10}$, $\xi_1=0.1$
\begin{figure}[H]
\vspace{-10pt}
     \centering
     \begin{subfigure}[b]{0.48\textwidth}
         \centering
         \includegraphics[width=\textwidth]{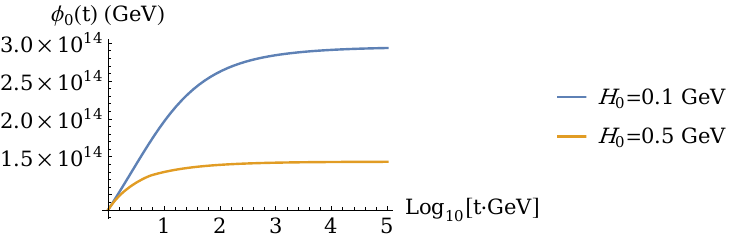}
     \end{subfigure}
     \hfill
     \begin{subfigure}[b]{0.48\textwidth}
         \centering
         \includegraphics[width=\textwidth]{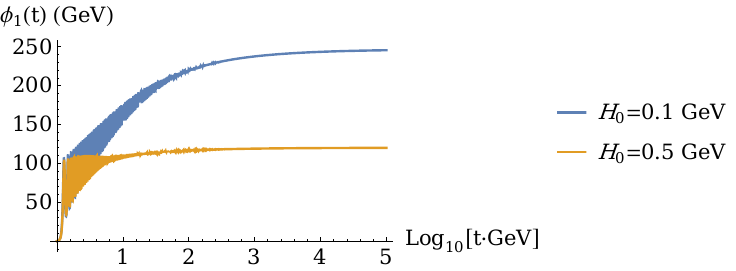}
    \end{subfigure}
    \\
     \begin{subfigure}[b]{0.48\textwidth}
         \centering
         \includegraphics[width=\textwidth]{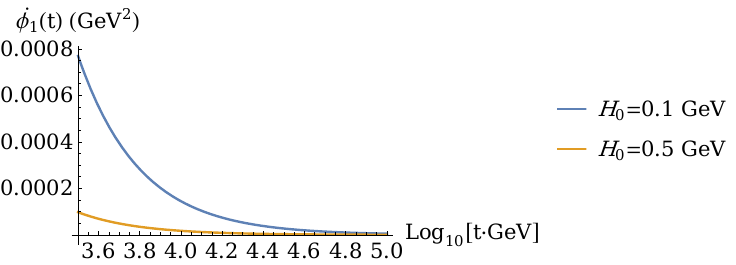}
    \end{subfigure}
    \hfill
    \begin{subfigure}[b]{0.48\textwidth}
         \centering
         \includegraphics[width=\textwidth]{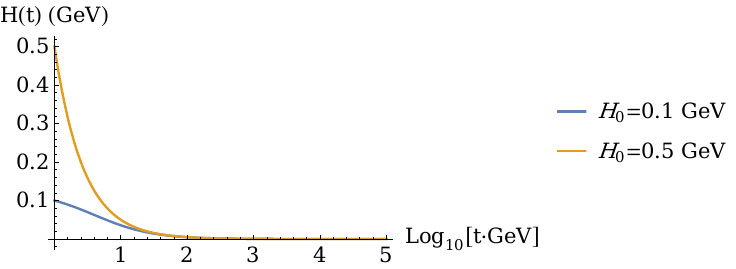}
     \end{subfigure}
     \\
     \begin{subfigure}[b]{0.48\textwidth}
         \centering
         \includegraphics[width=\textwidth]{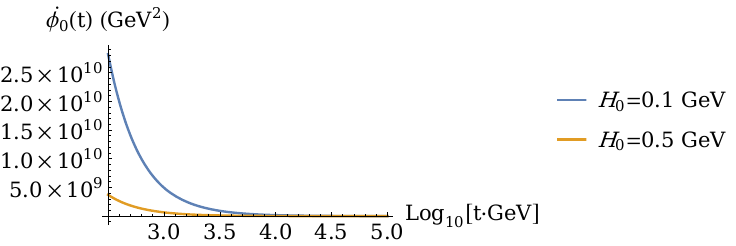}
    \end{subfigure}
    \hfill
    \begin{subfigure}[b]{0.48\textwidth}
         \centering
         \includegraphics[width=\textwidth]{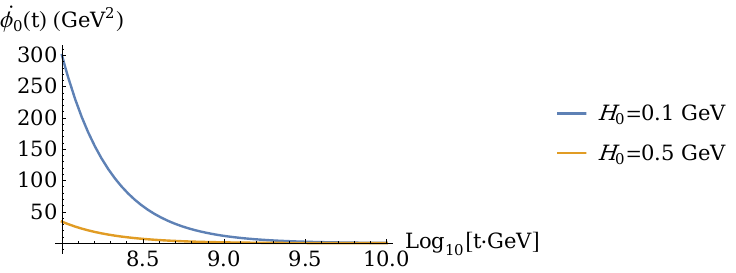}
     \end{subfigure}
    \caption{Evolution of $\phi_i$ fields and $H$ with time for coupling constants values fulfilling requirements from section \ref{Higgspotpar}: $\lambda_2=0.03125$, $\lambda_1=-4.37\cdot10^{-26}$, $\xi_0=10^{10}$, $\xi_1=0.1$. Initial conditions:  $\phi_0(0) = 8\cdot10^{13}$ GeV, $\dot{\phi_0}(0) = 5\cdot10^{13}$ GeV$^2$,  $\phi_1(0)=0$ GeV,   $\dot{\phi_1}(0)=10$ GeV$^2$, and two different $H(0)=H_0$ values. The bigger the initial $H_0$, the faster $\phi_i$ fields lose their velocity and settles in lower values. Two plots for $\dot{\phi_0}(t)$ are shown, one for the same time range as in the evolution of $\phi_1$ and $H$, one for later times, to show that $\phi_0$ indeed lose its velocity and settles in desired value.}
    \label{Evolutionplots_real}
    \vspace{-15pt}
\end{figure}

    \subsubsection*{Unrealistic model:} $\lambda_2=0.03125$, $\lambda_1=-10^{-6}$, $\xi_0=10^3$, $\xi_1=0.1$
    
\begin{figure}[H]
     \centering
     \begin{subfigure}[b]{0.48\textwidth}
         \centering
         \includegraphics[width=\textwidth]{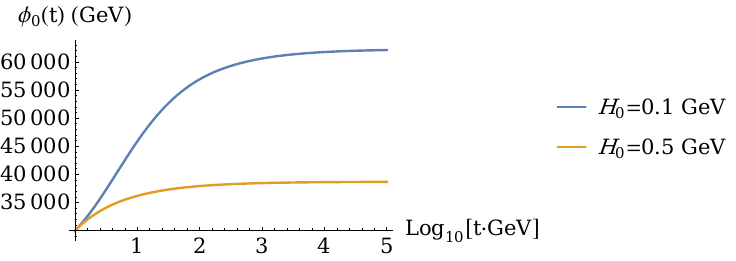}
     \end{subfigure}
     \begin{subfigure}[b]{0.48\textwidth}
         \centering
         \includegraphics[width=\textwidth]{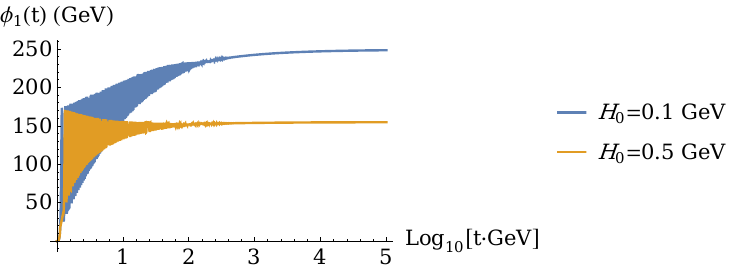}
    \end{subfigure}
    \\
    \begin{subfigure}[b]{0.48\textwidth}
         \centering
         \includegraphics[width=\textwidth]{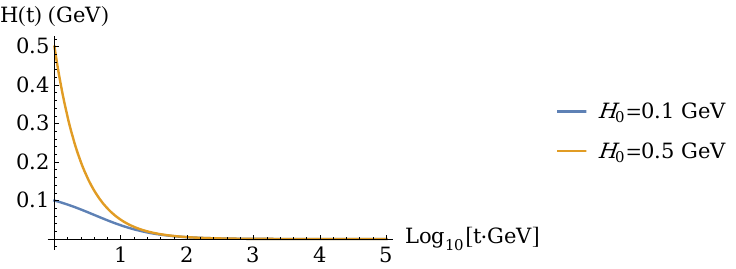}
     \end{subfigure}
     \begin{subfigure}[b]{0.48\textwidth}
         \centering
         \includegraphics[width=\textwidth]{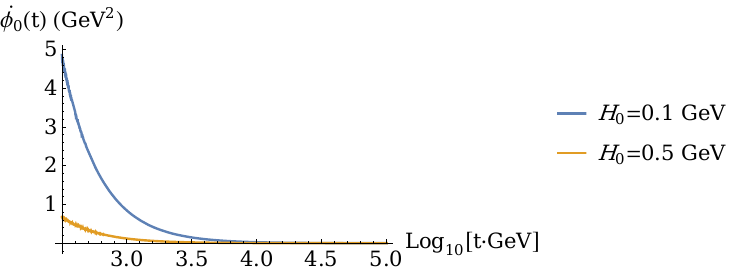}
     \end{subfigure}
    \caption{Evolution of $\phi_i$ fields and $H$ with time for coupling constants values: $\lambda_2=~0.03125$, $\lambda_1=-10^{-6}$, $\xi_0=10^3$, $\xi_1=0.1$. Initial conditions:  $\phi_0(0) = 3\cdot10^{4}$~GeV, $\dot{\phi_0}(0) = 5\cdot10^{3}$ GeV$^2$,  $\phi_1(0)=0$ GeV,   $\dot{\phi_1}(0)=10$ GeV$^2$ and two different $H(0)=H_0$ values. The bigger the initial $H_0$, the faster $\phi_i$ fields lose their velocity and settles in lower values.}
    \label{Evolutionplots_unreal}
\vspace{-15pt}
\end{figure}

\subsection{Non-zero temperature}

To examine evolution of fields $\phi_0$ and $\phi_1$ with time in a hot universe, we use equations of motion (\ref{EOMS}) corrected by thermal masses. Hence, we add to the potential $V$ the terms:
\begin{equation}
    V_{eff} =  V+\frac{1}{2}\phi_1^2\cdot\Big(\lambda_2+\frac{\lambda_1}{6}+\frac{g_1^2}{16}+\frac{3g_2^2}{16}+\frac{h_t^2}{4}\Big)T^2+\frac{1}{2}\phi_0^2\cdot\frac{\lambda_1}{6}T^2.
    \label{Tmodification}
\end{equation}

One could assume temperature dependence as in equation (\ref{H(T)dep}), which indicates the radiation dominated era. Unfortunately, using temperature $T$ as a function of $H$ and assuming domination of only radiation does not provide plausible conditions to obtain desired results of final evolution of the fields, i.e. Higgs vev of order 250 GeV. Therefore, we assume that Hubble parameter $H$ has distinct contribution from the $\phi_i$ fields and radiation contribution can be neglected in such case. Such system is immersed in temperature which dependence is ruled by radiation:
\begin{equation}
    T(t) = \frac{A}{\sqrt{\Big(t+t_0\Big)\cdot \textrm{GeV}}}, \qquad A = 1.6\cdot 10^9\textrm{ GeV},
    \label{T(t)}
\end{equation}
where $t_0$ is chosen to fit the initial eligible temperature of evolution $T_0 = 10^4$ GeV. We treat $H$ as an independent parameter ruled by the equations of motion (\ref{EOMSagain}) with temperature dependent potential (\ref{Tmodification}).

In general, the mechanism of kinetic energy dissipation via cosmic friction is sensitive to the chosen coupling value $\lambda_1$ and to the initial value of the Hubble parameter. If the chosen value of $\lambda_1$ is too large in magnitude, so the dilaton field is in thermal equilibrium, the temperature-dependent  contribution to the equation of motion for $\phi_0$ destabilizes the system and it can't evolve to the desired state with $\phi_1$ vacuum expectation value 250 GeV. Example evolution with $\lambda_1=-10^{-6}$ is shown on Figure \ref{Evolutionplots_temperature_Lambda1-6}.


\subsubsection*{Example evolution solutions}

To show differences in evolution with temperature, we choose the same initial conditions and parameter space values as in $T=0$ case. For $\lambda_1=-4.37\cdot10^{-26}$, $\phi_0$ field is not in thermal equilibrium, hence $\lambda_1\phi_0^2T^2$ term from (\ref{Tmodification}) is absent in equations of motion.

\begin{figure}[!h]
     \centering
     \begin{subfigure}[b]{0.48\textwidth}
         \centering
         \includegraphics[width=\textwidth]{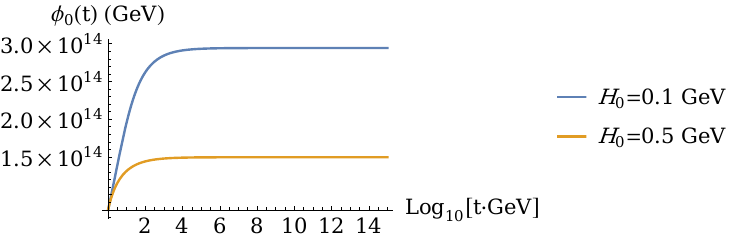}
     \end{subfigure}
     \hfill
     \begin{subfigure}[b]{0.48\textwidth}
         \centering
         \includegraphics[width=\textwidth]{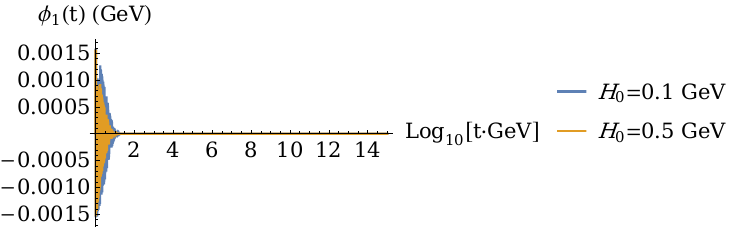}
    \end{subfigure}
    \\
    \begin{subfigure}[b]{0.48\textwidth}
         \centering
         \includegraphics[width=\textwidth]{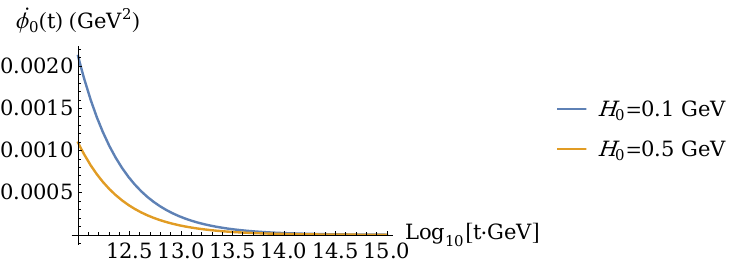}
     \end{subfigure}
     \hfill
     \begin{subfigure}[b]{0.48\textwidth}
         \centering
         \includegraphics[width=\textwidth]{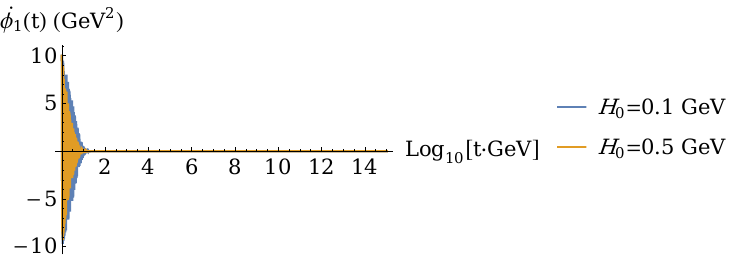}
    \end{subfigure}
     \\
    \begin{subfigure}[b]{0.48\textwidth}
         \centering
         \includegraphics[width=\textwidth]{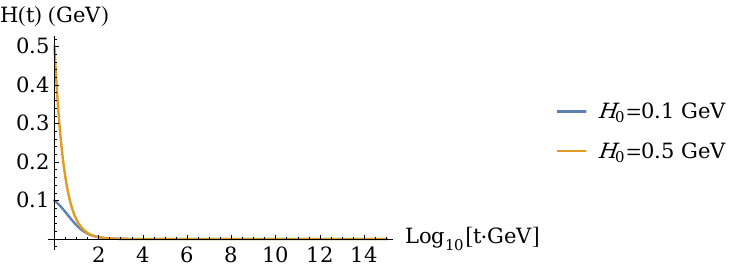}
     \end{subfigure}
     \hfill
     \begin{subfigure}[b]{0.48\textwidth}
         \centering
         \includegraphics[width=\textwidth]{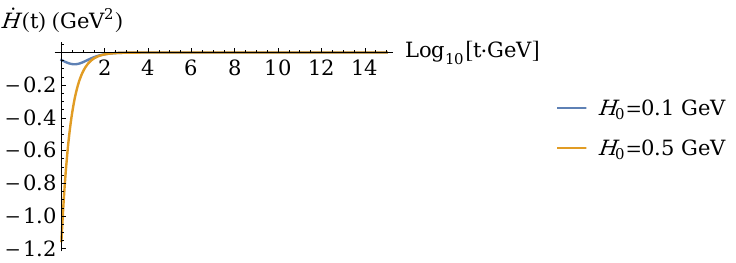}
    \end{subfigure}
    \caption{Evolution of $\phi_i$ fields and $H$ with time for non-zero temperature and coupling constants values fulfilling requirements from section \ref{Higgspotpar}: $\lambda_2=0.03125$, $\lambda_1=-4.37\cdot10^{-26}$, $\xi_0=10^{10}$, $\xi_1=0.1$. Initial conditions:  $\phi_0(0) = 8\cdot10^{13}$ GeV, $\dot{\phi_0}(0) = 5\cdot10^{13}$ GeV$^2$,  $\phi_1(0)=0$ GeV,   $\dot{\phi_1}(0)=10$ GeV$^2$ and two different $H(0)=H_0$ values. Initial temperature $T_0=10^4$ GeV.}
\label{Evolutionplots_real_temperature}
    \vspace{-10pt}
\end{figure}

\begin{figure}[H]
     \centering
     \begin{subfigure}[b]{0.46\textwidth}
         \centering
        \includegraphics[width=\textwidth]{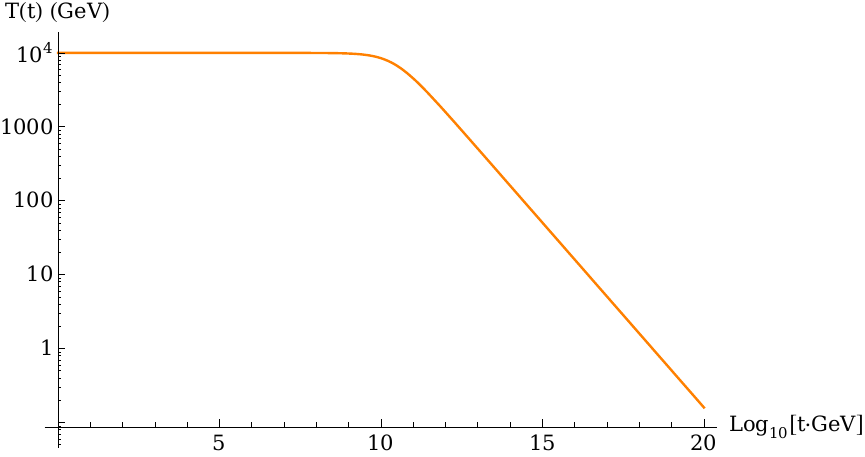}
         \caption{Temperature dependence (\ref{T(t)}) for $T_0~=~10^4$~GeV.}
         \label{Tt}
     \end{subfigure}
     \hfill
     \begin{subfigure}[b]{0.51\textwidth}
         \centering
    \includegraphics[width=\textwidth]{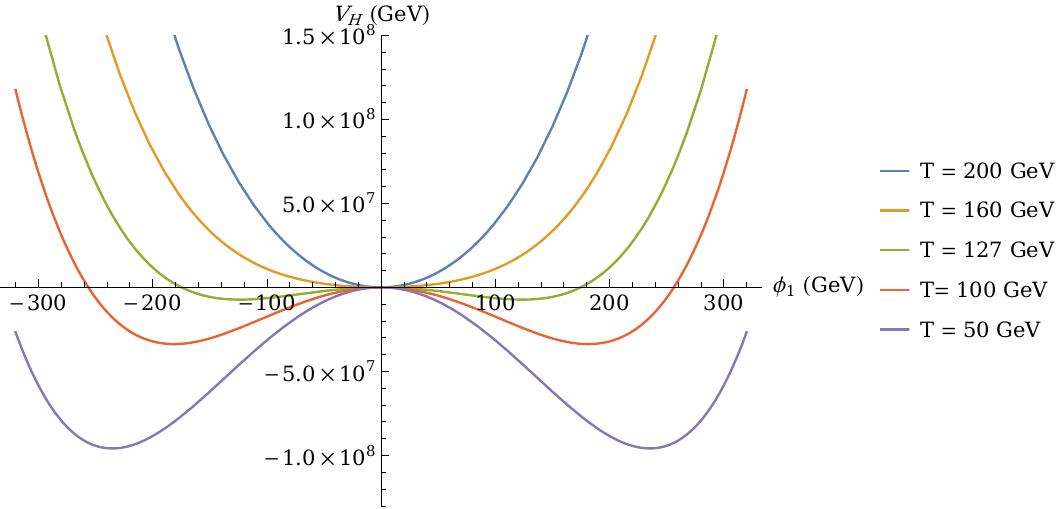}
         \caption{Higgs potential $V_H(\phi_1,T)$ for different temperatures in time.}
         \label{H26T}
    \end{subfigure}
    \caption{}
    \label{T+Higgs}
    \vspace{-10pt}
\end{figure}

It is easy to see from Figure \ref{Evolutionplots_real_temperature} that the dilaton field evolves to the same final value as in zero temperature case and after sufficiently long time $\dot{\phi_0}$ goes to zero. This is the case, because adding temperature to the potential with $\lambda_1\sim 10^{-26}$ doesn't affect the dilaton sector. However, the Higgs field is driven to zero. To understand this behaviour, on Figure~\ref{T+Higgs} we provide plots of temperature dependence (\ref{T(t)}) and Higgs potential $V_H$:
\begin{equation}
    V_H(\phi_1,T) = \lambda_0\bar{\phi_0}^4 + \lambda_1\bar{\phi_0}^2\phi_1^2+\lambda_2\phi_1^4 + \frac{1}{2}\phi_1^2 T^2\Bigg(\frac{\lambda_1}{6} + \lambda_2 + \frac{h_t^2}{4} + \frac{g_1^2}{16} +\frac{3g_2^2}{16}  \Bigg),
\end{equation}
where we fixed $\bar{\phi_0} = 3\cdot 10^{14}$ GeV, which corresponds to final $\phi_0$ value for $H_0=0.1$ GeV in non-zero temperature.

\newpage

    \subsubsection*{Realistic model, $\phi_0$ never in thermal equilibrium:} $\lambda_2=0.03125$, $\lambda_1=-4.37\cdot10^{-26}$, $\xi_0=10^{10}$, $\xi_1=0.1$

\begin{figure}[H]
    \centering
    \includegraphics[scale=0.81]{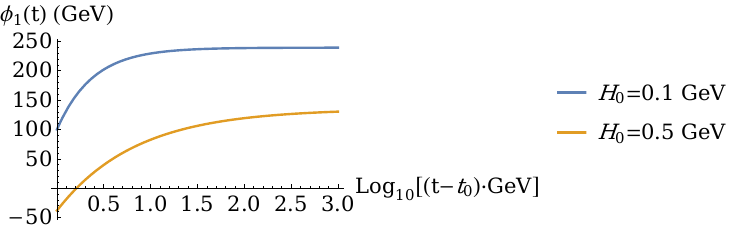}
    \caption{Further Higgs field evolution in time with initial values $\phi_1 = \delta\phi_1 = 37$ GeV and $\dot{\phi_1}=0$ GeV$^2$. Because $\phi_1$ oscillates around its minimum in this evolution, the plot shows $\phi_1$ average in time. $H_0$ values correspond to the initial conditions from Figure \ref{Evolutionplots_real_temperature}.}
    \label{Further_evolution}
\end{figure}

For temperatures higher than around 130 GeV, $V_H$ has only one minimum in zero value and because of that, Higgs is driven to this value during the evolution and stops. Since the potential in $\phi_1=0$ is flat and $\dot{\phi_1} = 0$ for later times, even if the temperature drops enough to produce two degenerate minima in $V_H$, Higgs field stays in origin in our simulation. But for fields in equilibrium in non-zero temperature, which is the case for $\phi_1$, there is associated a rms fluctuation \cite{Mukhanov} (see also high temperature limit of (\ref{Tpw})):
\begin{equation}
    \delta\phi_1 \approx \frac{T}{\sqrt{24}}.
\end{equation}
Hence we implement further evolution of the Higgs field with initial value for $\phi_1$ as this fluctuation, starting from time $t_0 = 10^{14.2}/$GeV which corresponds to sufficiently low temperature $T=127$ GeV for two degenerated minima to appear. Initial conditions for the rest of the parameters are their corresponding values for $t_0$ time. In this further evolution $\phi_0(t)$, $\dot{\phi_0}(t)$, $H(t)$ and $\dot{H}(t)$ don't change their values, hence we provide only the plot for $\phi_1(t)$ on Figure~\ref{Further_evolution}.

    \subsubsection*{Unrealistic model, $\phi_0$ in thermal equilibrium:} $\lambda_2=0.03125$, $\lambda_1=-10^{-6}$, $\xi_0=10^{3}$, $\xi_1=0.1$

\begin{figure}[H]
     \centering
     \begin{subfigure}[b]{0.48\textwidth}
         \centering
         \includegraphics[width=\textwidth]{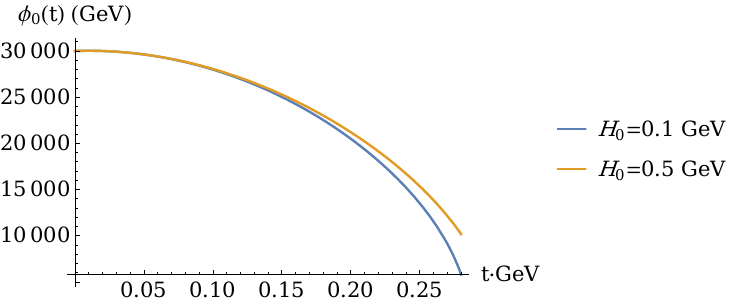}
     \end{subfigure}
     \hfill
     \begin{subfigure}[b]{0.48\textwidth}
         \centering
         \includegraphics[width=\textwidth]{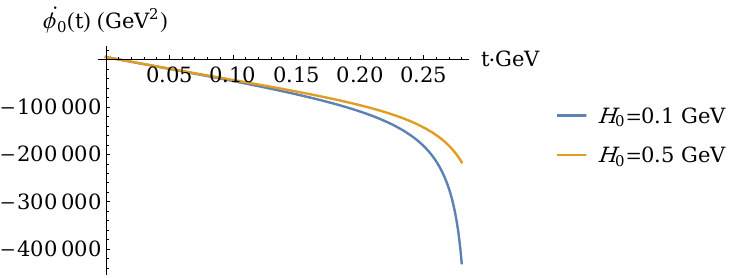}
    \end{subfigure}
    \\
     \begin{subfigure}[b]{0.48\textwidth}
         \centering
         \includegraphics[width=\textwidth]{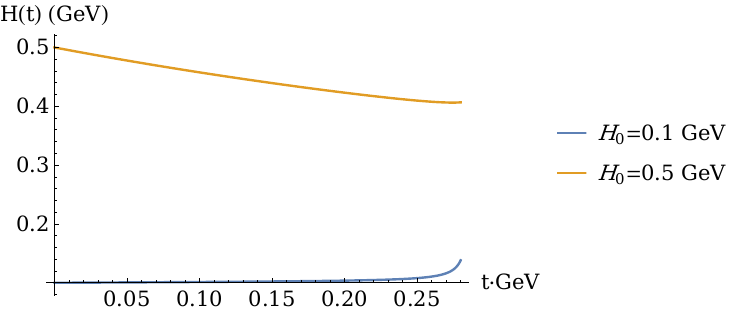}
    \end{subfigure}
    \hfill
    \begin{subfigure}[b]{0.48\textwidth}
         \centering
        \includegraphics[width=\textwidth]{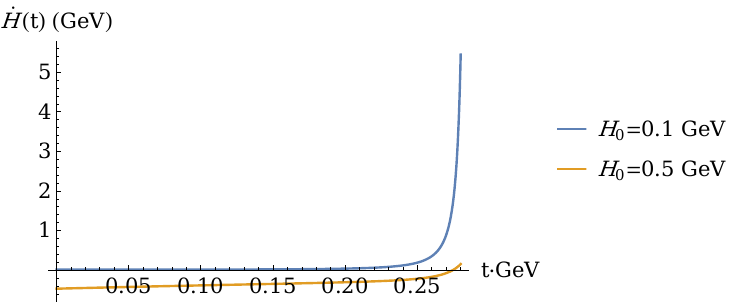}
     \end{subfigure}
    \caption{Evolution of $\phi_i$ fields and $H$ with time for non-zero temperature and coupling constants values $\lambda_2=0.03125$, $\lambda_1=-10^{-6}$, $\xi_0=10^{3}$, $\xi_1=0.1$. Initial conditions:  $\phi_0(0) = 3\cdot10^4$ GeV, $\dot{\phi_0}(0) = 5\cdot10^3$ GeV$^2$,  $\phi_1(0)=0$ GeV,   $\dot{\phi_1}(0)=10$ GeV$^2$ and two different $H(0)=H_0$ values. Initial temperature $T_0=10^4$ GeV. For larger $|\lambda_1|$ value, dilaton field is in thermal equilibrium and temperature contribution to its equation of motion makes the system unstable.}
\label{Evolutionplots_temperature_Lambda1-6}
\end{figure}

In this case, dilaton is in thermal equilibrium and $\lambda_1\phi_0^2 T^2$ term from (\ref{Tmodification}) is present in equations of motion. As it is shown on Figure \ref{Evolutionplots_temperature_Lambda1-6}, the system is unstable in this scenario. Higgs field oscillates around its zero value minimum for $T\neq 0$, same as on Figure \ref{Evolutionplots_real_temperature}, therefore we provide only $\phi_0$, $\dot{\phi_0}$, $H$ and $\dot{H}$ evolution plots on Figure \ref{Evolutionplots_temperature_Lambda1-6}.
    
\subsection{$T=0$ vs $T\neq0$}

The fact that the Universe is hot during expansion points out, that adding temperature to equations of motion is essential. Choosing parameters so dilaton is in thermal equilibrium, destabilizes the system. $\lambda_1\phi_0^2T^2$ term is then present not only in the equation of motion for $\phi_0$, but it also affects the gravity equation for the Hubble parameter $H$ (third equation in (\ref{EOMSagain})). Performed simulations indicate that only realistic scenario, with $|\lambda_1|$ not too large in magnitude, can provide sufficiently stable system, so it can evolve to desired mass scales values.

\subsection{Electroweak Symmetry Breaking}

\begin{figure}[!h]
\vspace{-0pt}
    \centering
    \includegraphics[width=0.9\textwidth]{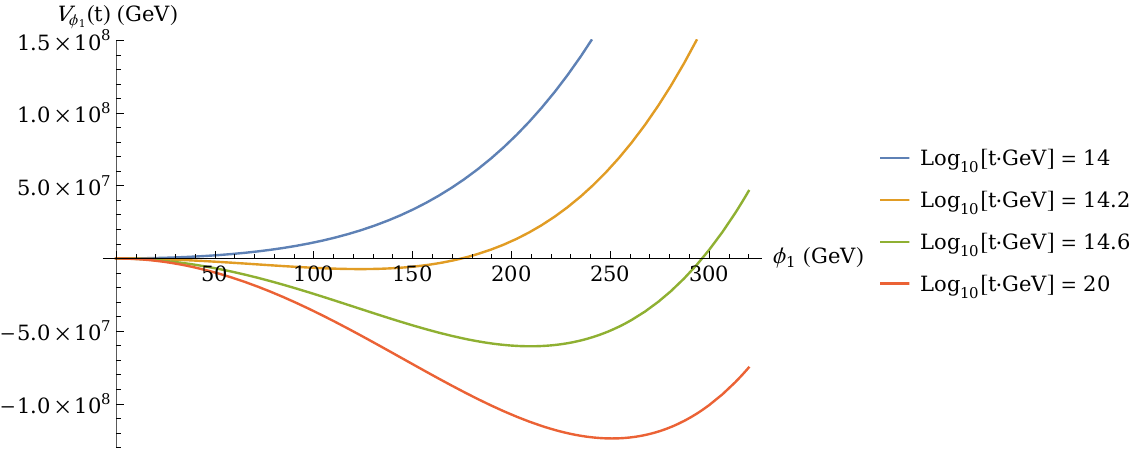}
    \caption{Plot of the potential for Higgs neutral component $V_{\phi_1}(t)$ for different time values during evolution in hot Universe.}
    \label{higgspot}
    \vspace{-10pt}
\end{figure}

To examine EWSB in our model, we consider time evolution of the effective potential $V_{eff}$ (\ref{Tmodification}) for the Higgs neutral component $\phi_1$, i.e.:
\begin{equation}
    V_{\phi_1}(t) \equiv V_{eff}\big(\phi_0(t),\phi_1,T(t)\big),
\end{equation}
where $\phi_0(t)$ comes from solution of evolution for realistic model parameters with $\lambda_2=~0.03125$, $\lambda_1 = -4.37\cdot 10^{-26}$ and time dependent temperature $T(t)$ (\ref{T(t)}). In Figure \ref{higgspot} we show plot of the shape of $V_{\phi_1}(t)$ for different time of evolution. It is easy to see that the phase transition associated with EWSB in tested model is of the second order. 

\section{Additional scalars}

Parameter space obtained from fulfilling numerical conditions on mass scales $m_H^2$, $v^2$ and $M_P^2$ is rather constrained. One can relax these constraints by adding more scalar singlets to the model. Here we consider briefly the addition of one more scalar singlet for simplicity. Then low energy Lagrangian density will take the form
\begin{equation}
    \frac{\mathcal{L}_{mod}}{\sqrt{g}} = -\frac{1}{12}\Big(\xi_0\phi_0^2+\xi_1\phi_1^2+\xi_2\phi_2^2\Big)R + \sum_{i=0}^2\frac{1}{2}\partial_{\mu}\phi_i\partial^{\mu}\phi_1-V_{mod} (\phi_0,\phi_1,\phi_2),
\end{equation}
where potential $V_{mod}$ is assumed to be:
\begin{equation}
    V_{mod}(\phi_0,\phi_1,\phi_2) = \lambda_0\phi_0^4+\lambda_1\phi_0^2\phi_1^2+\lambda_2\phi_1^4+\lambda_3\phi_1^2\phi_2^2+\lambda_4\phi_2^4,
\end{equation}
so the new scalar $\phi_2$ doesn't couple directly to the dilaton. Such a structure could be justified for instance by the locality of couplings in extra dimensions (space-time or internal). Taking $\lambda_2$ and $\lambda_4$ order one while $\lambda_1, \, \lambda_3 \, \ll 1$,  one can easily arrange for a 
ground state displaying a hierarchy of vevs:
\begin{equation}
    \langle\phi_0^2\rangle \gg \langle\phi_1^2\rangle \gg \langle\phi_2^2\rangle
\end{equation}
and zero cosmological constant condition:
\begin{equation}
    V_{mod} (\langle\phi_0\rangle,\langle\phi_2\rangle,\langle\phi_2\rangle) = 0,
\end{equation}
where the explicit solution reads:
\begin{equation}
    \langle\phi_1^2\rangle = \frac{2 \lambda _1 \lambda _4}{\lambda _3^2-4 \lambda _2 \lambda _4}\langle\phi_0^2\rangle, \qquad \langle\phi_2^2\rangle = -\frac{\lambda_3}{2\lambda_4}\langle\phi_1^2\rangle  = -\frac{ \lambda _1 \lambda _3}{\lambda _3^2-4 \lambda _2 \lambda _4} \langle\phi_0^2\rangle, \qquad \lambda_0 = \frac{\lambda _1^2 \lambda _4}{4 \lambda _2 \lambda _4-\lambda _3^2}.
\end{equation}
Then the effective Planck mass may become almost independent of the field playing the role of the Higgs and assuming the smallest vev:
\begin{equation}
    M_P^2 \sim  \xi_0\phi_0^2+\xi_1\phi_1^2.
\end{equation}
Such an extension of the scalar sector would allow building additional hierarchy of scales, while the thermal evolution for each pair of coupled scalars may follow approximately the scenario outlined above, this leading to the physically acceptable final state.  

\section{Summary and discussion}

In this note, we have analysed the possible thermal corrections to the cosmological evolution of the scale symmetric scalar sector extending the standard Higgs sector. All sources of the explicit breaking of scale invariance other than the temperature corrections have been neglected. The eventual hierarchy of generated mass scales relies on a hierarchy of small couplings, which stays perturbatively stable. We have visualized the effects of the breaking of the scale symmetry by the thermal corrections and argued, that the dynamical restoration of the scale symmetry at low energies and late times due to thermal corrections dragging the expectation value of the dilaton towards the origin can be avoided in realistic physical models. 
We have also demonstrated with specific examples that starting with the acceptable initial conditions, one can reach the physically relevant vacuum configuration as the result of the evolution of the scale symmetric scalar sector in the hot universe. 

\appendix
\section{Appendix}
\label{appendix}

\subsection{Weyl gravity}

We will show the basic properties of Weyl's conformal geometry \cite{Ghilencea2018_1, Ghilencea:2018_2, Ghilencea:2019, Ghilencea2019_WeylR^2inflation, Ghilencea2020_WeylvsPalatini, Ghilencea2021_SMinWeyl, Ghilencea2021_cosmological_evolution, Ghilencea:2022_non-metricity}
and its application to construct an action invariant under conformal Weyl transformations. Discussion in this section will lead us to very important conclusion: Einstein gravity is a low energy limit of Weyl gravity.

The main difference between Riemannian gravity and Weyl gravity is the presence of Weyl gauge vector field $\omega_{\mu}$. Weyl quantities will be denoted with a tilde and Riemannian
ones stay without a tilde. Connection $\Tilde{\Gamma}^{\rho}_{\mu\nu}$ takes the form:
\begin{equation}
    \Tilde{\Gamma}^{\rho}_{\mu\nu} = \Gamma^{\rho}_{\mu\nu}+\frac{q}{2}\Big[\delta^{\rho}_{\mu}\omega_{\nu}+\delta^{\rho}_{\nu}\omega_{\mu}-g_{\mu\nu}\omega^{\rho}\Big],
\end{equation}
where $q$ is the coupling of $\omega_{\mu}$ to the scalar field $\phi$ from the Weyl covariant derivative:
\begin{equation}
    \Tilde{D}_{\mu}\phi = \Big(\partial_{\mu}-\frac{q}{2}\omega_{\mu}\Big)\phi.
    \label{CovDev}
\end{equation}
The system is torsion free i.e. $\Tilde{\Gamma}^{\rho}_{\mu\nu}=\Tilde{\Gamma}^{\rho}_{\nu\mu}$. It is easy to see, that if we take the limit $\omega_{\mu}\rightarrow 0$, Riemannian geometry is recovered and $\Tilde{\Gamma}^{\rho}_{\mu\nu}\rightarrow \Gamma^{\rho}_{\mu\nu}$. The Weyl covariant derivative
of the metric, in contrast to Riemannian one, is different from zero:
\begin{equation}
    \Tilde{\nabla}_{\mu}g_{\alpha\beta} = -q\omega_{\mu}g_{\alpha\beta}.
\end{equation}
This is why in Weyl geometry parallel transport of a vector along a closed curve changes not only its direction (as in Einstein gravity case) but also its length. The Weyl curvature
scalar takes form:
\begin{equation}
    \Tilde{R} = R -3qD_{\mu}\omega^{\mu}-\frac{3}{2}q^2\omega_{\mu}\omega^{\mu},
\end{equation}
where $D_{\mu}$ is the Riemannian covariant derivative with the Levi-Civita connection $\Gamma^{\rho}_{\mu\nu}$. Weyl field
$\omega_{\mu}$ is abelian, thus the field strength tensor $\Tilde{F}_{\mu\nu}$ for $\omega_{\mu}$ is equal to the Riemannian one:
\begin{equation}
    \Tilde{F}_{\mu\nu} = \Tilde{D}_{\mu}\omega_{\nu}-\Tilde{D}_{\nu}\omega_{\mu} = \partial_{\mu}\omega_{\nu}-\partial_{\nu}\omega_{\mu} = F_{\mu\nu},
\end{equation}
where $\Tilde{D}_{\mu}$ is covariant derivative with connection $\Tilde{\Gamma}^{\rho}_{\mu\nu}$.

Weyl gauge invariance of the action is manifested by invariance under conformal transformations of the metric $g_{\mu\nu}$, scalar field $\phi$ and Weyl vector field $\omega_{\mu}$:
\begin{equation}
    \begin{split}
        g_{\mu\nu}\quad & \longrightarrow \quad g'_{\mu\nu} = \Omega^2g_{\mu\nu}, \\
        \phi \quad & \longrightarrow \quad \phi' = \frac{\phi}{\Omega}, \\
        \omega_{\mu} \quad & \longrightarrow \omega'_{\mu} = \omega_{\mu} - \frac{1}{q}\partial\ln\Omega^2,
    \end{split}
    \label{ConfTrans}
\end{equation}
where $\Omega$ is dimensionless parameter. Then $\sqrt{g'} =\Omega^4\sqrt{g}$, $\Tilde{R}' = \Omega^{-2}\Tilde{R}$ and $\Tilde{F}'_{\mu\nu} = \Tilde{F}_{\mu\nu}$.

\subsection{Stueckelberg mechanism for Weyl vector field}

Now we are going to consider the theory containing only Weyl quadratic gravity and vector field $\omega_{\mu}$ in the absence of matter. The original Lagrangian for Weyl quadratic gravity invariant under (\ref{ConfTrans}) can be of the form:
\begin{equation}
   \frac{\mathcal{L_W}}{\sqrt{g}} = \frac{\xi_0}{4!}\Tilde{R}^2-\frac{1}{4}g^{\mu\nu}g^{\rho\sigma}\Tilde{F}_{\mu\rho}\Tilde{F}_{\rho\sigma},
\end{equation}
where $\xi_0>0$. The $\Tilde{R}^2$ term propagates additional scalar state, so we can substitute:
\begin{equation}
    \Tilde{R}^2 \quad \rightarrow \quad -2\phi_0^2\Tilde{R}-\phi_0^4.
\end{equation}
Equation of motion for $\phi_0$ gives $\phi_0^2 = -\Tilde{R}$, thus the Weyl quadratic gravity term is recovered and we can consider an equivalent Lagrangian of the form:
\begin{equation}
    \frac{\mathcal{L_W}}{\sqrt{g}} = -\frac{\xi_0}{12}\phi_0^2\Tilde{R}-\frac{1}{4}g^{\mu\nu}g^{\rho\sigma}\Tilde{F}_{\mu\rho}\Tilde{F}_{\rho\sigma}-\frac{\xi_0}{4!}\phi_0^4.
    \label{LWwithfi0}
\end{equation}
In the above Lagrangian we have one massless scalar field $\phi_0$ and one massless vector field $\omega_{\mu}$ present in $\Tilde{R}$ and $\Tilde{F}_{\mu\nu}$ terms. Now, we apply (\ref{ConfTrans}) transformations to it with $\Omega = \xi_0\phi_0^2/6M^2$:
\begin{equation}
    \phi_0'^2 = \frac{\phi_0^2}{\Omega} = \frac{6M^2}{\xi_0}, \qquad \Tilde{R}' = R'-\frac{3}{2}q^2\omega'_{\mu}\omega'^{\mu}+(\textrm{total deriv.})
\end{equation}
Neglecting the total derivative term, we obtain:
\begin{equation}
    \frac{\mathcal{L'_W}}{\sqrt{g'}} = -\frac{1}{2}M^2R'+\frac{3}{4}q^2M^2\omega'_{\mu}\omega'^{\mu}-\frac{1}{4}\Tilde{F}'_{\mu\nu}\Tilde{F}'^{\mu\nu}-\frac{3}{2\xi_0}M^4,
    \label{EPaction}
\end{equation}
where $R'$ is the Riemannian curvature scalar after transformation (\ref{ConfTrans}). Weyl vector field became massive via the Stueckelberg mechanism $m_{\omega}^2 = \frac{3}{4}q^2M^2$. Usually $M$ is chosen to~$M_{Planck}$. So Planck mass scale is naturally generated after breaking of the Weyl conformal symmetry (via Stueckelberg mechanism). These phenomena, impossible in Riemannian gravity, can give the solution to the hierarchy problem and originates mass scales in physics.

The Lagrangian (\ref{EPaction}) is the Einstein-Proca action for Weyl gauge field. It is worth observing that the number of degrees of freedom is conserved. We had massless vector
and a scalar field, which gives 3 d.o.f. and after the gauge symmetry transformation we have only one vector massive field, still 3 d.o.f. This is happening because the vector field $\omega_{\mu}$ has “eaten” dilaton mode $\phi_0$ (actually the real dilaton is $\ln\phi_0^2$ with shift symmetry:
$\ln\phi_0^2\rightarrow\ln\phi_0^2-\ln\Omega$, but we refer to $\phi_0$ field as dilaton in this work, since it remains massless). Without $\omega_{\mu}$ conservation of d.o.f. wouldn't be possible. In $\mathcal{L_W}$ there is also a positive cosmological constant term.
For a coupling $q$ not too small, the mass of the Weyl vector field is close to the Planck's mass. Below this scale, the $\omega_{\mu}$ decouples and the whole theory becomes Riemannian. So Einstein gravity can be considered as the low energy limit of Weyl gravity. 

\subsubsection{Matter fields}

In this paper, we are considering a scale symmetric extension of the Standard Model Higgs scalar sector. Therefore, in addition to Lagrangian (\ref{LWwithfi0}), we want to consider a more general case. Let us denote $\phi_1$ as a Higgs neutral component. Lagrangian invariant under (\ref{ConfTrans}) is:
\begin{equation}
    \frac{\mathcal{L}}{\sqrt{g}} = -\frac{1}{12}\Big(\xi_0\phi_0^2+\xi_1\phi_1^2\Big)\Tilde{R}-\frac{1}{4}\Tilde{F}_{\mu\nu}\Tilde{F}^{\mu\nu}+\frac{1}{2}\Tilde{D}_{\mu}\phi_0\Tilde{D}^{\mu}\phi_0+\frac{1}{2}\Tilde{D}_{\mu}\phi_1\Tilde{D}^{\mu}\phi_1-V(\phi_0,\phi_1),
    \label{LWplusHiggs}
\end{equation}
where we added kinetic terms for $\phi_i$ fields with covariant derivative (\ref{CovDev}) and $V(\phi_0,\phi_1)$ is scale symmetric potential, which includes possible interactions between $\phi_i$ fields and will be specified later.

Now, we can apply the transformations (\ref{ConfTrans}) to (\ref{LWplusHiggs}) with:
\begin{equation}
    \Omega = \frac{\xi_0\phi_0^2+\xi_1\phi_1^2}{6M^2}.
\end{equation}
One gets:
\begin{equation}
    \frac{\mathcal{L}'}{\sqrt{g'}} = -\frac{1}{2}M^2R'+\frac{3}{4}q^2M^2\omega'_{\mu}\omega'^{\mu}-\frac{1}{4}\Tilde{F}'_{\mu\nu}\Tilde{F}'^{\mu\nu}+\frac{1}{2}\Tilde{D}'_{\mu}\phi_0'\Tilde{D}'^{\mu}\phi_0'+\frac{1}{2}\Tilde{D}'_{\mu}\phi_1'\Tilde{D}'^{\mu}\phi_1'-V'(\phi_0',\phi_1'),
\end{equation}
which again is the Einstein-Proca action with Weyl gauge field and two $\phi_0$, $\phi_1$ scalars. Initially massless $\omega_{\mu}$ field, acquires mass $m_{\omega}^2 = \frac{3}{4}q^2M^2$ and for energies below that scale, can be considered as decoupled and the whole theory becomes Riemannian.

The above discussion shows that after spontaneous scale symmetry breaking, i.e. when scalar fields $\phi_i$ acquire their vevs, so that
\begin{equation}
    \frac{1}{6}\Big(\xi_0\langle\phi_0^2\rangle+\xi_1\langle\phi_1^2\rangle\Big) = M^2,
    \label{MPlanck}
\end{equation}
the mass scales are generated. 


\section*{Acknowledgments}
This work has been supported by the Polish National Science Center grant 2017/27/B/ST2/02531.

\end{document}